\RequirePackage{luatex85}
\documentclass[12pt]{article}
\usepackage{tikz}
\usepackage{amssymb}
\usepackage{epsfig}
\usepackage{bbm}
\usepackage{bbold}
\usepackage{graphicx}
\usepackage{hyperref}
\usepackage{romannum}
\usepackage{graphics,graphpap}
\usepackage{amsmath,exscale,amsthm}
\usepackage{yfonts}
\usepackage{mathrsfs}
\usepackage{float}
\usepackage{hyperref}
\usepackage{subcaption}
\usepackage{axodraw}
\usepackage[compat=1.1.0]{tikz-feynman} 
 \usepackage{feynmf}
 \usepackage{slashed}

\usepackage[none]{hyphenat}
\usetikzlibrary{matrix,positioning,decorations.pathreplacing}

\usepackage{color}
\usepackage{amssymb}
\usepackage{epsfig}
\usepackage{romannum}

\def\e{\varepsilon}

\newcommand{\wt}{\widetilde}

\usepackage{amsmath}
\usepackage{amsfonts}

\begin{document}

\def\a{\alpha}
\def\b{\beta}
\def\c{\chi}
\def\d{\delta}
\def\e{\epsilon}
\def\f{\phi}
\def\g{\gamma}
\def\h{\eta}
\def\i{\iota}
\def\j{\psi}
\def\k{\kappa}
\def\la{\lambda}
\def\m{\mu}
\def\n{\nu}
\def\o{\omega}
\def\p{\pi}
\def\q{\theta}
\def\r{\rho}
\def\s{\sigma}
\def\t{\tau}
\def\u{\upsilon}
\def\x{\xi}
\def\z{\zeta}
\def\D{\Delta}
\def\F{\Phi}
\def\G{\Gamma}
\def\J{\Psi}
\def\L{\Lambda}
\def\O{\Omega}
\def\P{\Pi}
\def\Q{\Theta}
\def\S{\Sigma}
\def\U{\Upsilon}
\def\X{\Xi}

\def\ve{\varepsilon}
\def\vf{\varphi}
\def\vr{\varrho}
\def\vs{\varsigma}
\def\vq{\vartheta}

\def\dg{\dagger}                                     
\def\ddg{\ddagger}                                   
\def\wt#1{\widetilde{#1}}                    
\def\mt{\widetilde{m}_1}
\def\mti{\widetilde{m}_i}
\def\rt{\widetilde{r}_1}
\def\mtt{\widetilde{m}_2}
\def\mttt{\widetilde{m}_3}
\def\rtt{\widetilde{r}_2}
\def\mb{\overline{m}}
\def\VEV#1{\left\langle #1\right\rangle}        
\def\be{\begin{equation}}
\def\ee{\end{equation}}
\def\ds{\displaystyle}
\def\ra{\rightarrow}

\def\bea{\begin{eqnarray}}
\def\eea{\end{eqnarray}}
\def\NO{\nonumber}
\def\Bar#1{\overline{#1}}


\def\pl#1#2#3{Phys.~Lett.~{\bf B {#1}} ({#2}) #3}
\def\np#1#2#3{Nucl.~Phys.~{\bf B {#1}} ({#2}) #3}
\def\prl#1#2#3{Phys.~Rev.~Lett.~{\bf #1} ({#2}) #3}
\def\pr#1#2#3{Phys.~Rev.~{\bf D {#1}} ({#2}) #3}
\def\zp#1#2#3{Z.~Phys.~{\bf C {#1}} ({#2}) #3}
\def\cqg#1#2#3{Class.~and Quantum Grav.~{\bf {#1}} ({#2}) #3}
\def\cmp#1#2#3{Commun.~Math.~Phys.~{\bf {#1}} ({#2}) #3}
\def\jmp#1#2#3{J.~Math.~Phys.~{\bf {#1}} ({#2}) #3}
\def\ap#1#2#3{Ann.~of Phys.~{\bf {#1}} ({#2}) #3}
\def\prep#1#2#3{Phys.~Rep.~{\bf {#1}C} ({#2}) #3}
\def\ptp#1#2#3{Progr.~Theor.~Phys.~{\bf {#1}} ({#2}) #3}
\def\ijmp#1#2#3{Int.~J.~Mod.~Phys.~{\bf A {#1}} ({#2}) #3}
\def\mpl#1#2#3{Mod.~Phys.~Lett.~{\bf A {#1}} ({#2}) #3}
\def\nc#1#2#3{Nuovo Cim.~{\bf {#1}} ({#2}) #3}
\def\ibid#1#2#3{{\it ibid.}~{\bf {#1}} ({#2}) #3}

\title{
\vspace*{1mm}
{\bf Completing RHINO}

\author{
{\Large Pasquale Di Bari and  Adam Murphy}
\\
{\it Physics and Astronomy}, 
{\it University of Southampton,} \\
{\it  Southampton, SO17 1BJ, U.K.} 
\\
}}
\maketitle \thispagestyle{empty}
\pagenumbering{arabic}

\begin{abstract}
The right-handed (RH)  Higgs-induced neutrino mixing  (RHINO) model explains 
neutrino masses and origin of matter in the universe within a unified picture.
The mixing, effectively described by a dimension five operator, 
is responsible both for the production of dark neutrinos, converting a small fraction
of seesaw neutrinos acting as source, and for their decays. 
We show that including the production of source neutrinos from Higgs portal  interactions, 
their abundance can thermalise prior to the onset of source-dark neutrino oscillations,
resulting into an enhanced production of dark neutrinos that thus
can play the role of decaying dark matter (DM) for a much higher seesaw scale. 
This can be above the sphaleron freeze-out temperature and as high as $\sim 100\,{\rm TeV}$, 
so that strong thermal resonant leptogenesis for the generation of the matter-antimatter asymmetry is viable.  
We obtain a $\sim 1\,{\rm TeV}$--$1\,{\rm PeV}$ allowed dark neutrino mass range. 
Intriguingly, their decays can also  explain a  neutrino flux excess 
at ${\cal O}(100\,{\rm TeV})$ energies recently confirmed by the IceCube collaboration analysing 7.5yr HESE data.  
Our results also point to an effective scale for Higgs portal interactions nicely identifiable with 
the grandunified scale and many orders of magnitude below the effective scale for the mixing.
We explain this  hierarchy  in a UV-complete model with a very heavy fermion as mediator:
the first scale corresponds to the  fundamental scale of new physics, while the second is much higher because of a very small coupling 
that can be identified with a symmetry breaking parameter. Therefore, RHINO realises a simple unified model of neutrino masses 
and origin of matter  in the universe currently under scrutiny at neutrino telescopes and potentially embeddable within a grandunified model.  
\end{abstract}

\newpage

\section{Introduction}

The origin of matter in the universe is a key issue in modern physics \cite{origin}. Its solution would guide us along 
that path to new physics whose existence is  motivated  both to a theoretical and a phenomenological level.  A strong phenomenological motivation is represented by the necessity to extend  the standard model (SM) in order to incorporate neutrino masses and mixing. It is then reasonable to investigate whether extensions of the SM able to explain neutrino masses and mixing  can also provide a solution to the problem of the origin of matter in the universe. The minimal type-I seesaw mechanism \cite{seesaw}, 
augmenting the SM particle content with the introduction of RH neutrinos, 
provides an elegant and minimal way to understand both the lightness of the ordinary neutrinos 
and the observed large mixing angles in the leptonic mixing matrix. Moreover, it can be easily embedded 
within many realistic models beyond the SM \cite{King:2015aea} and
it leads to leptogenesis as a built-in scenario of baryogenesis for the explanation of the matter-antimatter asymmetry of the universe \cite{fy}.  It is also well known that a keV seesaw RH neutrino can be efficiently produced via active-sterile neutrino mixing and play the role of (warm) DM \cite{dw}. A unified picture, combining a keV lightest seesaw neutrino DM produced from active-sterile neutrino mixing
with successful leptogenesis from sterile-sterile neutrino mixing \cite{ars} and satisfying 
low energy neutrino data is also potentially viable \cite{nuMSM}. However,
constraints from X-ray observations and large scale structure N-body simulations require the presence of a large lepton asymmetry for an efficient resonant production \cite{shifuller}, introducing an additional issue. In principle, 
such a large lepton asymmetry can  be produced by the decays of one  of the two heaviest seesaw neutrinos,  
without additional non-minimal ingredients beyond the type-I seesaw Lagrangian \cite{canetti}, but
this requires a very strong degeneracy of the two heavier RH neutrino masses. 
Interestingly, a 7 keV lightest seesaw neutrino resonantly produced and playing the role of DM can also address the 3.5 keV X-ray anomaly \cite{abazajian}. Recently, numerical solutions fulfilling all requirements and the 3.5 keV anomaly have been shown to exist \cite{ghiglieri}.  
However, the relative degeneracy of the two heavier RH neutrino masses has to be very fine-tuned, at the level of $10^{-16}$. 
In any case the XRISM satellite \cite{XRISM} should soon either confirm the 3.5 keV anomaly or place stringent constraints on warm DM decays,
ruling out this minimal scenario for the origin of matter in the universe 
where new physics resides just in the type-I seesaw Lagrangian.

An alternative approach, where a RH neutrino plays the role of cold rather than warm DM, 
is based on adding a nonrenormalizable operator  to the type-I seesaw Lagrangian \cite{ad}.
In this case a {\em dark neutrino} $N_{\rm D}$, with suppressed Yukawa couplings, would be produced by its Higgs-induced 
mixing with a seesaw neutrino $N_{\rm S}$ via the dimension five Anisimov operator \cite{ad,anisimov}
\be\label{anisimov}
{\cal L}^{\rm A}_{\rm DS} = {\la_{\rm DS} \over \L} \, \Phi^\dagger \, \Phi \, \overline{N_{\rm D}^c} \, N_{\rm S} \,  ,
\ee
where $\Lambda$ is the scale of new physics. 
This mixing would be both responsible for the production of the dark neutrinos and for 
their decays \cite{ad,unified}. Decays would be indeed induced at zero temperature 
by the effective small Yukawa couplings generated by the mixing \cite{densitym}. 
In this way there is a tension between the two conditions to be imposed for the dark neutrino to be 
a DM candidate:  reproducing the observed DM abundance and being sufficiently long-lived to evade constraints 
from  neutrino telescopes \cite{gelminisarkar}. This tension produces constraints on the neutrino mixing effective scale $\widetilde{\L}_{\rm DS} \equiv \L/\la_{\rm DS}$ and on the mass of the dark neutrino \cite{densitym}, making the scenario highly predictive.

The production of dark neutrinos proceeds through their Higgs-induced mixing with the seesaw neutrinos $N_{\rm S}$ that act as a source 
and, for this reason, we refer to them as {\em source neutrinos}. The origin of the mixing is  a mismatch between
the Majorana mass eigenbasis and the basis where the new Higgs-induced interactions are diagonal. It  has a strong temperature dependence
encoded in the thermal self-energy that, in the case of neutrino mixing, can be described in terms of the effective potential
or, alternatively, in terms of thermal masses. This strong temperature dependence is the key for a solution of the
tension between an efficient production and a sufficiently long-lived dark neutrino, 
since the mixing can be much stronger in the very early universe than today, when dark neutrinos decay. 

Initially, calculations of the dark neutrino abundance were done using a Landau-Zener approximation \cite{ad,unified}. However, it has been shown that,
in general, this fails to provide a correct description and numerical calculations within a density matrix formalism show 
that the production is, in general, much less efficient \cite{densitym}.\footnote{The validity of the 
Landau-Zener approximation requires extremely quasi-degenerate dark-source neutrino masses.}
If one starts from a vanishing source neutrino abundance, a stringent upper bound is obtained on the mass of the 
source neutrino mass that needs to be lighter than the W gauge boson mass. In this way the four-body decay rate gets 
suppressed and the dark neutrino can be the DM particle only for masses $M_{\rm D} \gtrsim 10^7\,{\rm GeV}$. 
However, in such a setup, strong thermal (i.e., independent of the initial conditions) leptogenesis from (non-relativistic) decays is not viable, since
the matter-antimatter asymmetry should be  resonantly produced from the decays of the seesaw neutrinos with masses 
about twice the sphaleron freeze-out temperature $T_{\rm RH} \sim 132\,{\rm GeV}$ \cite{sphaleron}. 
The asymmetry could be still generated from the mixing of the two seesaw neutrinos with GeV masses \cite{ars,nuMSM} or also
from decays \cite{Drewes:2021nqr,Granelli:2020ysj}. Therefore, compared to the $\nu$MSM,  one would still have low scale leptogenesis with 
GeV seesaw neutrinos but the DM would be very heavy and, therefore, cold rather than warm. 

However, in low scale leptogenesis with GeV neutrinos the final asymmetry is produced in the relativistic regime and it would be sensitive
not only to an accurate description of many different processes but also to the initial conditions. Since the reheat temperature
needs to be higher than $\sim 10^9\,{\rm GeV}$ for an efficient dark neutrino production, in this case a large pre-existing asymmetry is a rather 
natural outcome in many different mechanisms. Therefore, having to impose 
that this is negligible is certainly an unattractive feature of the scenario.
 In this paper we show that a rather simple and elegant solution to this problem exists without spoiling minimality. 

The main point is that the Anisimov operator is actually only one term of a more general effective
Lagrangian that can be written as \cite{ad} ($I,J = {\rm D, S}$)
\be\label{angenprime}
{\cal L}_{A} = \sum_{I,J} {\la_{\rm IJ}' \over \L} \, \Phi^\dagger \, \Phi \, \overline{N_{\rm I}^c} \, N_{\rm J} \,  ,
\ee
or, explicitly, as the sum of three terms 
\be\label{angen}
{\cal L}_{A} =  {\la_{\rm DS} \over \L} \, \Phi^\dagger \, \Phi \, \overline{N_{\rm D}^c} \, N_{\rm S}
+ {\la_{\rm SS} \over \L} \, \Phi^\dagger \, \Phi \, \overline{N_{\rm S}^c} \, N_{\rm S} 
+ {\la_{\rm DD} \over \L} \, \Phi^\dagger \, \Phi \, \overline{N_{\rm D}^c} \, N_{\rm D} \,  ,
\ee
having properly redefined $\lambda_{\rm DS} = \lambda_{\rm DS}'+\lambda_{\rm SD}'$.\footnote{This is because 
one simply has $\overline{N_{\rm S}^c} \, N_{\rm D} = \overline{N_{\rm D}^c} \, N_{\rm S}$.}
In this way also (nonrenormalisable) Higgs portal interactions terms are included.\footnote{Renormalisable 
Higgs portal interactions to scalars were  first considered in \cite{Patt:2006fw}, 
while nonrenormalisable 5-dim Higgs portal interactions to RH neutrinos in \cite{ad}.} 
Higgs portal interactions were  assumed not to play a role in the production of dark neutrinos and they were neglected in 
previous papers \cite{ad,unified,densitym}.  
Here we show that actually these interactions can influence the production of dark neutrinos in a very significant way
and help not only to reconcile a solution of the DM conundrum with strong thermal  leptogenesis
but also to allow dark neutrino to be a decaying DM for masses $M_{\rm D}\sim 100\,{\rm TeV}$, 
currently hinted by measurements of high energy neutrino flux at neutrino telescopes.  
Notice that in Eq.~(\ref{angen}) there is also a term describing Higgs portal interactions for dark neutrinos that 
might compete with the mixing term  in the production of dark neutrinos. The same mixing term can also be  responsible 
for a mixed production of dark neutrinos and source neutrinos from Higgs scatterings ($\phi+\phi^\dagger \ra N_{\rm D}+N_{\rm S}$).
However, in this paper we assume that the production of dark neutrinos from Higgs scatterings 
can be neglected. In the end of the paper we will discuss a model where this assumption is naturally justified.
Notice, however, that direct dark neutrino production from Higgs scatterings might 
play a role in other contexts, for example in the thermalisation of a dark sector 
prior to a phase transition, as in the scenario discussed in \cite{DiBari:2021dri}.

In this paper we show that including Higgs portal interactions for the source neutrinos, 
the allowed region in the space of parameters enlarges at higher seesaw scales, where traditional leptogenesis from decays 
in the non-relativistic regime can work independently of the initial conditions (strong thermal leptogenesis). 
This is because they can partly or fully thermalise the source neutrinos prior to the mixing, i.e., at higher temperatures than Yukawa interactions. This makes possible to reproduce the observed DM abundance for higher values of the  effective RHINO scale  
$\widetilde{\Lambda}_{\rm DS} \equiv \Lambda/\lambda_{\rm DS}$ 
and, consequently, to satisfy  the constraint from four body decays also for values of the seesaw scale
above the W gauge boson mass. We show that in this way the seesaw scale can be as high as $\sim 100\,{\rm TeV}$. 
This is still  well below the  leptogenesis lower bound $M_I \gtrsim 10^9\,{\rm GeV}$ holding for hierarchical 
seesaw neutrinos \cite{di} so that a resonant production \cite{resonant1,resonant2}
and/or some combined tuning in the seesaw formula \cite{lepbounds,antuschblanchet,jturner} 
are necessary to enhance the (total and/or flavoured) $C\!P$ asymmetries.  
Moreover, such higher seesaw scales are attractive also because they can be more 
easily embedded within realistic models of flavour \cite{King:2015aea}. Moreover,  
for such higher seesaw scale, the allowed dark neutrino mass is in the range $\sim 1\,{\rm TeV}$--$1\,{\rm PeV}$,
thus making possible to address the hint of a $\sim 100\,{\rm TeV}$ excess in high energy neutrino data.

The paper is structured as follows. In Section 2 we briefly review the RHINO model and 
how the dark neutrino abundance  can be calculated within a density matrix formalism. In Section 3 we include
Higgs portal interactions for the source neutrinos showing how the set of kinetic equations gets modified 
and discuss some benchmark cases clearly illustrating their effect. 
Here we also show how the critical value of the effective scale for the thermalisation of source  neutrinos is nicely
obtained to be coinciding with the grandunified scale. In Section 4 we derive the allowed
regions in the dark neutrino lifetime-mass plane, for various values of the seesaw scale, reheat temperature and effective 
scale for the source neutrino Higgs portal interactions. 
In Section 5 we discuss a UV-complete model that can consistently incorporate all interactions described by the Anisimov operator.
This is able to explain the values of the three  effective scales as stemming from just one fundamental scale that can be nicely identified with 
the grandunified scale, something that should be regarded as a successful outcome of the model.  
Finally, in Section 6 we  draw our conclusions. 

\section{The RHINO model}

Let us now briefly review the RHINO model and how the dark neutrino abundance can be calculated within 
a density matrix formalism just using the Anisimov operator in Eq.~(\ref{anisimov}). 
The SM field content is augmented by the introduction of three RH neutrinos.   
However, in addition to the usual seesaw Lagrangian with neutrino Yukawa couplings and Majorana mass terms, 
one also has Higgs-induced neutrino mixing terms, so that ($I=J=1,2,3; \a=e,\mu,\tau$)
\be\label{lagrangian}
-{\cal L}^\n_{Y+M+\L} = \overline{L_{\a}}\,h_{\a I}\, N_{I}\, \widetilde{\Phi} +
                          \frac{1}{2} \, \overline{N^{c}_{I}} \, M_I \, N_{I}  +  
                          \sum_{I\neq J}{\la_{I J} \over \L} \, \Phi^\dagger \, \Phi \, \overline{N_{\rm I}^c} \, N_{J}
                          + \mbox{\rm h.c.}  \,  ,
\ee
where the $L_\a$'s are the three lepton doublets, $\Phi$ is the Higgs doublet and $\widetilde{\Phi} \equiv {\rm i}\,\sigma_2\,\Phi^{\star}$ is its dual.  
Notice here we are not (yet) including Higgs portal interactions corresponding to terms $I=J$.  
The neutrino Yukawa matrix  $h$ is written in a basis where both charged lepton mass matrix and Majorana mass matrix $M_{IJ}$ are diagonal so that the $N_I$'s are the three Majorana mass eigenstates. One of the three RH neutrinos is assumed to have vanishing Yukawa couplings and, therefore,
it is fully decoupled from the seesaw formula.\footnote{Therefore, at this stage, we should actually 
refer to $N_{\rm D}$ as a heavy neutral lepton. However, small Yukawa couplings are triggered by
the Anisimov operator, as we are going to discuss soon.}   
Moreover, we assume that the Higgs-induced mixing between the two seesaw neutrinos is negligible, we will comment on this assumption. 
In our discussion we identify the dark neutrino $N_{\rm D}$ with the heaviest RH neutrino $N_3$,
so that its mass $M_{\rm D}=M_3$. This corresponds to assume neutrino Yukawa matrix of the form
\be\label{yukawa}
 h =
 \left( \begin{array}{ccc}
h_{e 1}    &  h_{e 2}  & 0  \\
h_{\m 1} &  h_{\m 2} & 0 \\
h_{\t 1}  &  h_{\t 2} & 0
\end{array}\right) \,   .
\ee
This matrix can be diagonalised by a bi-unitary transformation of the left-handed and RH neutrino fields, i.e.,
\be
h = V_L^\dagger \, D_h \, U_R \,  ,
\ee
where $D_h \equiv {\rm diag}(h_C, h_B, h_A)$ with $h_A \leq h_B \leq h_C$.
Necessarily,  starting from a Yukawa matrix of the form (\ref{yukawa}), one has $h_A = 0$.    
The RH neutrino mixing matrix $U_R$ is of the form
\be
U_R(\ve) =  \left( \begin{array}{ccc}
 \cos \ve  &  \sin \ve & 0   \\
 -\sin \ve &  \cos  \ve & 0 \\
0  &  0 & 1
\end{array}\right) \,   ,
\ee
where $\ve$ is a complex angle that generates a mixing between the seesaw neutrinos $N_1$ and $N_2$. 
This mixing is necessary if one wants to have non-vanishing $C\!P$ asymmetries in $N_1$ and $N_2$-decays 
in order to generate a baryon asymmetry via leptogenesis. In our case, the most conservative choice, maximising 
the dark neutrino lifetime, is to take $\varepsilon \simeq 0$ so that approximately $U_R \simeq I$. 
Notice that even such an infinitesimal deviation of $U_R$ from the identity can yield successful leptogenesis \cite{Buchmuller:2002rq}.
Within this setup, ordinary neutrino masses and mixing are generated by a minimal two-RH neutrino
type I seesaw mechanism, where the two seesaw neutrinos have to be identified with $N_1$ and $N_2$ 
and the lightest neutrino mass vanishes. 

In order to maximise the  dark neutrino lifetime, 
we can also assume the dark neutrino mixing only with the seesaw neutrino with smaller 
Yukawa coupling $h_B$. This plays the role of source neutrino, denoted by $N_{\rm S}$,
and it can be either the lightest RH neutrino $N_1$ or the next-to-lightest $N_2$. 
For definiteness, we can choose $N_{\rm S} = N_2$ so that $M_{\rm S} = M_2$. 
For this reason in the following we will refer to a two-neutrino mixing formalism. 
In order to minimise the Yukawa coupling of the source neutrino
and maximise the lifetime of DM, we can also  assume normal ordering for the light neutrino masses.   
In this way one simply has  $h_{\rm S} =  h_B= {\sqrt{m_{\rm sol}\,M_{\rm S}}}/v$.
Of course, notice that with this choice one also has $h_C = {\sqrt{m_{\rm atm}\,M_{I}}}/v$,
where either $I=1$, if $N_{\rm S}=N_2$, or $I=2$, if $N_{\rm S} = N_1$. 
However, in order to have successful  leptogenesis at a scale much below $10^9\,{\rm GeV}$, necessarily $M_1 \simeq M_2$, 
so whether the source RH neutrino corresponds to the lightest or to the next-to-lightest seesaw neutrino does not 
make any difference in our discussion.  In conclusion, the Lagrangian in Eq.~(\ref{lagrangian}) simplifies to ($I=1,2,3$, $J=1,2$):
\be\label{lagrangian2}
-{\cal L}^\n_{Y+M+\widetilde{\L}_{\rm DS}} = \overline{L_{\a}}\,h_{\a J}\, N_{J}\, \widetilde{\Phi} +
                          \frac{1}{2} \, \overline{N^{c}_{I}} \, M_I \, N_{I}  +  
                          {1 \over \widetilde{\L}_{\rm DS}} \, \Phi^\dagger \, \Phi \, \overline{N_{\rm D}^c} \, N_{{\rm S}}
                          + \mbox{\rm h.c.}  \,  .
\ee
After electroweak spontaneous symmetry breaking, the Higgs acquires a vev  $v$ that generates a neutrino
Dirac mass matrix 
\be
m_D = v\,h = \left( \begin{array}{ccc}
m_{D\, e 1}    &  m_{D\, e 2}  & 0  \\
m_{D\,\m 1} &  m_{D\,\m 2} & 0 \\
m_{D\, \t 1}  &  m_{D\,\t 2} & 0
\end{array}\right) \,   
\ee
from the usual Yukawa term but also a small off-diagonal correction to the Majorana 
mass term $\d M^\L _{\rm DS}=  v^2/\widetilde{\L}_{\rm DS}$ generated by the Anisimov operator. 
Therefore, the full neutrino mass Lagrangian is now given by
\be
-{\cal L}_{D+M+\d M^{\L}_{\rm DS}}^{\nu}= \overline{\nu_{L\a}}\,m_{D\a J}\, N_{J} + 
{1\over 2}\,\overline{N^{c}_{I}}\,M_I \, N_{I} + 
                          {1 \over 2} \, \overline{N_{{\rm D}}^c} \, \delta M^{\L}_{{\rm D S}} \, N_{{\rm S}} + {\rm h.c.}
 \,  .
\ee
In this way, as an effect of the Anisimov operator, 
the mass eigenstates change from  $N_{\rm D}, N_{\rm S}$ to
\bea
N^\Lambda_{{\rm D} 0} & = & N_{\rm DM} \cos\theta_{\L 0}  - N_{\rm S}\,\sin\theta_{\L 0}  \\
N^\Lambda_{{\rm S} 0} & = & N_{\rm DM} \, \sin\theta_{\L 0}  + N_{\rm S} \, \cos\theta_{\L 0} \,  ,
\eea
where\footnote{This expression holds for $\theta_{\L 0}\ll 1$. This is always verified unless one considers 
a case of extreme degeneracy, in which case the angle becomes maximal. Our results are not at all affected 
by such an approximation.}
\be\label{thetaL0}
\theta_{\L 0} = 
\frac{2\, v^2/\widetilde{\L}_{\rm DS}}{M_{\rm D}\,(1-M_{\rm S}/M_{\rm D})} 
\ee
is the effective mixing angle at zero temperature. In this way the RH neutrino mixing matrix from 
the new mass eigenstate basis to the flavour basis can be written as
\be
U_R^\Lambda(\theta_\Lambda^0) = \left( \begin{array}{ccc}
 1  &  0 & 0   \\
 0 &  \cos \theta_\Lambda^0 & \sin\theta_\Lambda^0 \\
0  &  -\sin\theta_\Lambda^0 & \cos \theta_\Lambda^0
\end{array}\right)
\ee
and in the new basis the neutrino Dirac mass matrix becomes:
\be
m^\Lambda_D = m_D \, U_R^\Lambda(\theta_\Lambda^0)  = 
\left( \begin{array}{ccc}
m_{D\, e 1}    &  m_{D\, e 2}\,\cos \theta_\Lambda^0  & m_{D\, e 3}\,\sin \theta_\Lambda^0  \\
m_{D\,\m 1} &  m_{D\,\m 2}\,\cos \theta_\Lambda^0 & m_{D\,\m 3}\,\sin \theta_\Lambda^0 \\
m_{D\, \t 1}  &  m_{D\,\t 2}\,\cos \theta_\Lambda^0 &  m_{D\,\t 3}\,\sin \theta_\Lambda^0
\end{array}\right) \,  .
\ee
This shows that now the mass eigenstates created by the field $N^\Lambda_{\rm D}$, those that 
have to be identified with the DM particles,  will not be rigorously stable but decay with a lifetime \cite{densitym} 
\be\label{lifetimebound}
\tau_{\rm D} \simeq (\G_{{\rm D} \ra A + \ell_{\rm S}}  + \G_{{\rm D} \ra 3A + {\ell}_{\rm S}})^{-1} \,  ,
\ee
where $\G_{{\rm D} \ra A + \ell_{\rm S}}$ is the two body decay rate given by
\be
\Gamma_{{\rm D} \ra A + \ell_{\rm S}}  = 
{h^2_{\rm S} \over \pi} \, \left(\frac{v^2}{\widetilde{\L}}\right)^2
\, {M_{\rm D}\over (M_{\rm D} - M_{\rm S})^2}\,  
\ee
and $\G_{{\rm D} \ra 3A + {\ell}_{\rm S}} $ is the four body decay rate given, in the narrow width
approximation, by
\be
\G_{{\rm D} \ra 3A + {\ell}_{\rm S}} = 
{\Gamma_{\rm S} \over 15 \cdot 2^{11} \cdot \pi^{4}} \, {M_{\rm D} \over M_{\rm S}} \, 
\left({M_{\rm D} \over \widetilde{\L}_{\rm DS}}\right)^2 \,  ,
\ee
where $\G_{\rm S} = h^2_{\rm S}\,M_{\rm S}/(4\,\pi)$.  The most stringent lower bound on the lifetime  comes from the
IceCube collaboration that, analysing 7.5 yr HESE data, has recently found approximately $\tau_{\rm D}\gtrsim 10^{27}$--$10^{28}\,{\rm s}$ 
in the energy range $60\,{\rm TeV}$--$10\,{\rm PeV}$ \cite{IceCube:2022vtr},  
extending previous results \cite{IceCube:2018tkk}.\footnote{At the moment, this should be regarded
as an indicative approximate lower bound. We will later on discuss more detailed  lower bounds 
that in fact depends on the DM mass and on the decay channel.}
 
At finite temperatures, the Yukawa and Anisimov interactions in Eq.~(\ref{anisimov})
both contribute to the self-energies of the RH neutrinos affecting their propagation and mixing.  
These are diagrammatically shown in panels (a) and (b) of Fig.~1.

\begin{figure}[t]
\centerline{\psfig{file=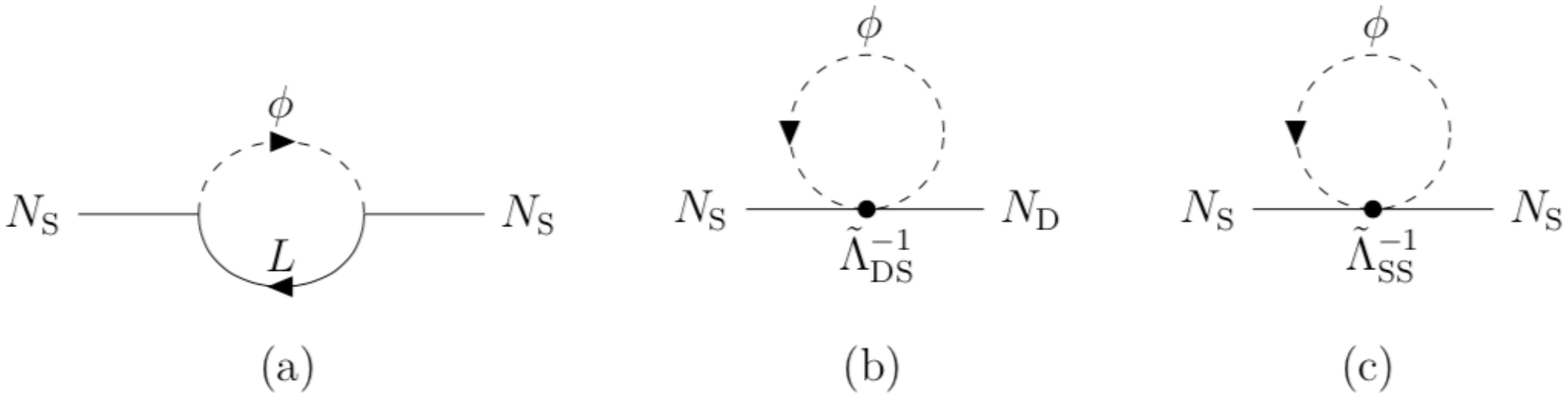,height=15cm,width=11cm,angle=-90}}\vspace{-35mm}
\caption{Self-energy diagrams from Yukawa interactions (panel (a)) and from Anisimov interactions (panels  (b) and (c)).}
\end{figure}

The effect can be described in terms of the effective potentials that, in the Yukawa basis, are given respectively by \cite{weldon}
($I,J={\rm D},{\rm S}$) 
\be\label{yukawapot}
V^{Y}_{IJ} =  \frac{T^2}{8\,E_J} \, h^2_J \, \d_{IJ}  \;  ,
\ee
and
\be\label{LAMBDAmixing}
V^{\L}_{I J} \simeq \frac{T^2}{12\,\L}\,\la_{I J} \,  .
\ee  
The contribution from the Anisimov interactions is not diagonal in general in the Yukawa basis and this misalignment generates
a mixing that strongly depends on temperature.  The evolution of the dark and source neutrino eigenstates
will be then described by the Hamiltonian
\be\label{hamiltonian}
{\cal H} =  \left( \begin{array}{cc}
E_{\rm DM} &  \frac{T^2}{12\,\widetilde{\L}_{\rm DS}} \\[1ex]
\frac{T^2}{12\,\widetilde{\L}_{\rm DS}}  &  E_{\rm S} + \frac{T^2}{8\,E_{\rm S}} \, h^2_{\rm S}
\end{array}\right)  \,  .
\ee
Subtracting a term proportional to the identity, the effective mixing Hamiltonian is then given by
\be\label{effectiveham}
\Delta {\cal H} \simeq   
\left( \begin{array}{cc}
- \frac{\D M^2}{4 \, p} - \frac{T^2}{16\,p} \, h^2_{\rm S} &  \frac{T^2}{12\,\widetilde{\L}_{\rm DS}}  \\[1ex]
\frac{T^2}{12\,\widetilde{\L}_{\rm DS}} &  \frac{\D M^2}{4 \, p} + \frac{T^2}{16 \, p} \, h^2_{\rm S}  
\end{array}\right)  \, ,
\ee
where we used the ultrarelativistic approximation\footnote{This assumes that the dark neutrino production
occurs in the ultra-relativistic regime.} and 
defined $\D M^2 \equiv M^2_{\rm S} - M^2_{\rm D}$. 

The Yukawa interactions produce a source neutrino abundance that we denote by $N_{N_{\rm S}}$
and normalise it in a way that in ultrarelativistic thermal equilibrium $N_{N_{\rm S}}^{\rm eq}(z_{\rm S} \ll 1) = 1$,
where $z_{\rm S} \equiv M_{\rm S}/T = z\,M_{\rm S} /M_{\rm D}$ and $z \equiv M_{\rm D}/T$.
The production of source neutrinos is then described by the simple rate equation
\be\label{rateequation}
\frac{dN_{N_{\rm S}}}{dz} = - (D +S)\,(N_{N_{\rm S}}-N_{N_{\rm S}}^{\rm eq})  \,  ,
\ee
where  $D \equiv \G_D/(H\,z)$, $S \equiv \G_{\rm S}/(H\,z)$ and $\G_{\rm D}$  and $\G_{S}$ 
are the source neutrino total decay  and the $\D L = 1$ scattering rates respectively. 
Finally, $H=H(z)$ is the expansion rate given by
\be
H(z) = \sqrt{8\pi^3\,g_R \over 90} \, {M_{\rm D}^2\over M_{\rm P}} \, {1 \over z^2} \, ,
\ee
where for the value of the number of ultrarelativistic degrees of freedom  we can simply take
the SM value, so that $g_R =g_R^{\rm SM} = 106.75$.  
The off-diagonal term in the Hamiltonian will then mix source and dark neutrinos.
Adopting a monochromatic approximation,  the momentum can be replaced by its average value $p \simeq 3\,T$ 
and the effective mixing Hamiltonian in the flavour basis becomes
\be
\Delta {\cal H} \simeq   
\frac{\D M^2}{12\,T}\,\left( \begin{array}{cc}
	- 1 - v_{\rm S}^Y &  \sin 2\theta_{\L}  \\[1ex]
\sin 2\theta_{\L}   &  1 + v_{\rm S}^Y 
\end{array}\right)  \,  .
\ee
Here  we have also introduced the dimensionless effective potential $v_{\rm S}^Y \equiv T^2\,h^2_{\rm S} / (4\,\D M^2)$ and the effective mixing angle $\sin 2\theta_{\L}(T) \equiv T^3/(\widetilde{\L}_{\rm DS} \, \D M^2 )$, 
parameterising the misalignment between the Yukawa and the Higgs-induced interactions.

The production of dark neutrinos can be described by the density matrix equation \cite{densitym}
\be\label{densitymatrixeq}
{d {{\cal N}} \over dz} = -{i\over H(z)z}\,[\D{\cal H}, {\cal N}]  - 
\begin{pmatrix}
0   &  {1\over 2}(D+S) \,{\cal N}_{\rm DS}  \\ 
{1\over 2}(D+S)  \,{\cal N}_{\rm SD}  & (D+S)\,(N_{N_{\rm S}}-N_{N_{\rm S}}^{\rm eq})  
\end{pmatrix} \,   ,
\ee
where the diagonal elements give the abundances of dark neutrinos, $N_{N_{\rm D}} = {\cal N}_{\rm DD}$, and
source neutrinos, $N_{N_{\rm S}} = {\cal N}_{\rm SS}$. Notice how decays and scatterings also contribute
to decoherence effects, damping the density matrix off-diagonal terms.
Initially, for $z=z_{\rm in}$, the density matrix is simply given by:
\be
{\cal N}(z_{\rm in}) = N_{N_{\rm S}}(z_{\rm in})\,\left(\begin{array}{cc}
0 & 0 \\
0 & 1
\end{array}\right) \,  .
\ee 
Expressing the matrices in the Pauli matrix basis, the density matrix equation 
can be recast in a vectorial notation. The effective Hamiltonian can be written as 
\be\label{DH}
\D{\cal H} = {1\over 2}\,\vec{V} \cdot \vec{\s}  \,  , 
\ee
where the {\em effective potential vector} $\vec{V}$ is defined as
\be
\vec{V} \equiv \frac{\D M^2}{6\,T}\,\left(\sin 2\theta_{\L} , 0 , - 1 - v_{\rm S}^Y \right)  \,   .
\ee 
The abundance normalised density matrix is analogously recast, introducing the 
quantity $P_0$ and the polarisation vector $\vec{P}$, as 
\be\label{N}
{\cal N} =  {1\over 2}\,P_0 \, \left( 1 + \vec{P} \cdot \vec{\s} \right)\,  ,
\ee
in a way that 
\bea
N_{N_{\rm D}}  &  =  &  {1\over 2}\,P_0 \, \left( 1 + P_z \right) \,   , \\ 
N_{N_{\rm S}}     &  =  &  {1\over 2}\,P_0 \, \left( 1 - P_z \right) \,  ,\\  
N_{N_{\rm D}} + N_{N_{\rm S}} & = &  P_0 \,  . 
\eea
Inserting Eqs.~(\ref{DH}) and (\ref{N}) into the density matrix equation (\ref{densitymatrixeq}),
one obtains a set of equations for  $P_0$ and $\vec{P}$:
\bea\label{densitymatrixeqpauli}
{d  \vec{P} \over d z}& = & \vec{\overline{V}} \times \vec{P} -
\left[{1\over 2}(D+S)  + {d\ln P_0 \over dz} \right]\,\vec{P}_T - 
(1 + P_z)\,{d \ln P_0 \over d z} \,\hat{z} \,   , \\ \label{production}
{d  P_0 \over d z} & = & - (D+S)\,(N_{N_{\rm S}}-N_{N_{\rm S}}^{\rm eq})  \,   ,
\eea
where we defined $\vec{P}_T \equiv P_x\,\hat{x} + P_y \, \hat{y}$ and
$\vec{\overline{V}} \equiv \vec{V}/(H\,z)$.

In Fig.~2 we show the evolution of $N_{N_{\rm D}}$ and $N_{N_{\rm S}}$ for the benchmark values
$M_{\rm S} = 300\,{\rm GeV}$, $M_{\rm D} = 220\,{\rm TeV}$, $\tau_{\rm D} = 3.48 \times 10^{28}\,{\rm s}$
and assuming an initial vanishing $N_{\rm S}$ abundance. 
These particular values for $M_{\rm D}$ and $\tau_{\rm D}$ correspond to best fit values found in a likelihood statistical analysis 
of 6 year High Energy Starting Events (HESE) IceCube where the presence of 
a neutrino flux contribution from neutrinophilic DM decays in addition to  an astrophysical 
component with fixed spectral index $\gamma =2.2$ is favoured at more than $\sim 3\sigma$ compared to
the null hypothesis where there is  no decaying DM \cite{Chianese:2018ijk}.\footnote{Another analysis of
6 year HESE data where, differently from \cite{Chianese:2018ijk}, the spectral index is let to vary and DM decays via one decay channel, also finds that the
addition of a component from decaying DM improves the fit. The best fit is obtained for DM decaying into W bosons, a mass $M_{\rm DM} \sim 400\,{\rm TeV}$ and for a value of the spectral index $\gamma \simeq 2.3$ but with a lower  $\sim 2\sigma$ statistical significance \cite{Bhattacharya:2019ucd}.
Also in a more recent 7.5 yr HESE data analysis a maximum $\sim 2\sigma$ statistical significance is found in various decay channels \cite{Chianese:2019kyl}.}

\begin{figure}[t]
\centerline{\psfig{file=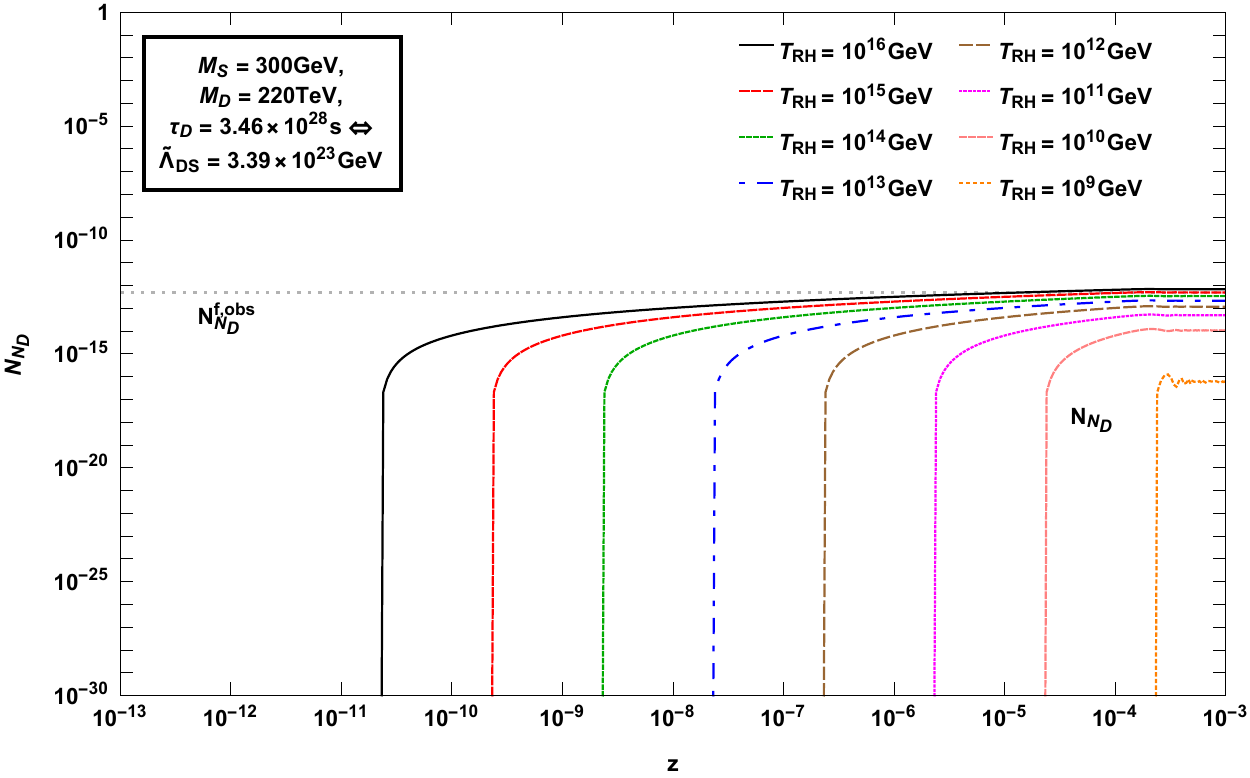,height=6.3cm,width=10cm,angle=0}}
\vspace{5mm}
\centerline{\psfig{file=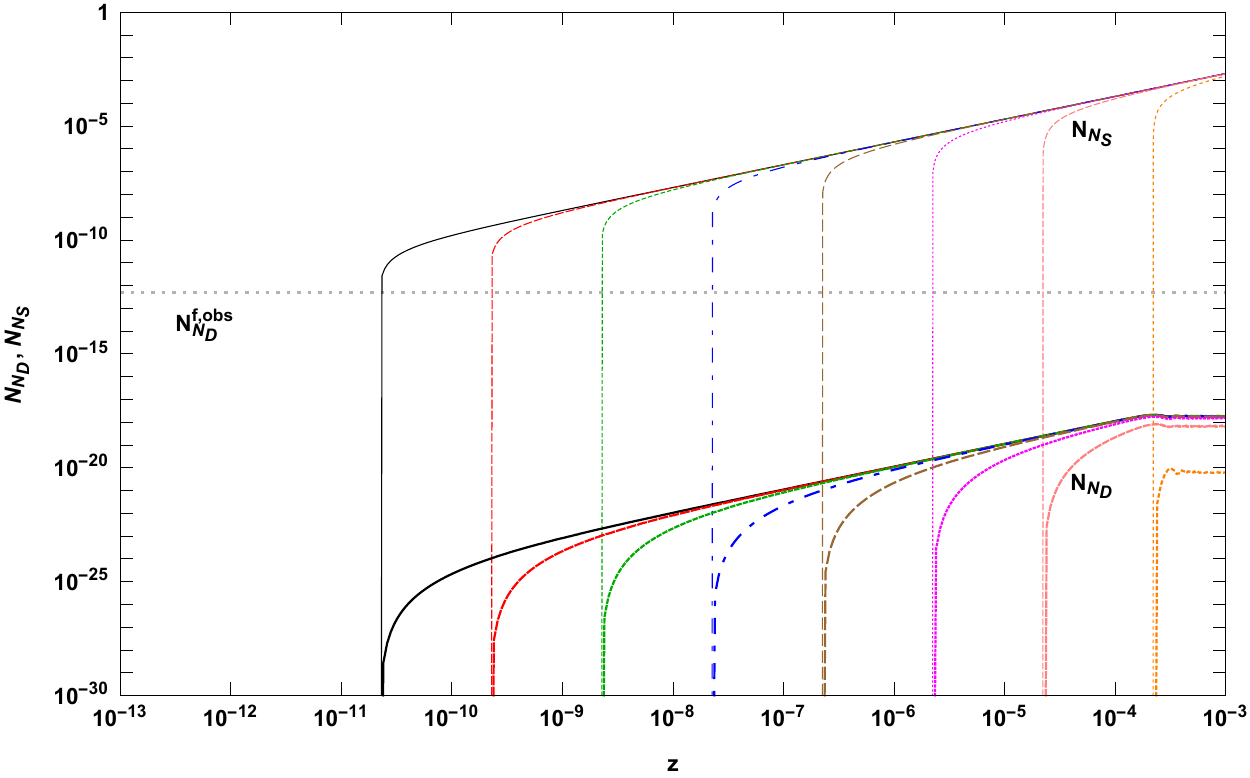,height=6.3cm,width=10cm,angle=0}}
\caption{Evolution of the source and dark neutrino abundances for different values of the initial temperature and for the indicated benchmark values of $M_{\rm D}, M_{\rm S}$ and $\tau_{\rm D}$. Top panel: initial thermal $N_{N_{\rm S}}$-abundance. Bottom panel: initial vanishing $N_{N_{\rm S}}$-abundance.}
\end{figure}
The horizontal line denotes $N_{N_{\rm D}}^{\rm f,obs}$, the final value of $N_{N_{\rm D}}$ that reproduces the measured DM energy 
density parameter at the present time $\Omega_{\rm DM}\, h^2 \simeq 0.119$ \cite{planck18}. 
For a generic mass $M_{\rm D}$, one finds \cite{densitym}
\be\label{NDobs}
N_{N_{\rm D}}^{\rm f,obs} \simeq 1.1 \times 10^{-7} \, {{\rm GeV} \over M_{\rm D}} \,  .
\ee
One can notice how the final dark neutrino abundance falls many orders of magnitude below $N_{N_{\rm D}}^{\rm f, obs}$. 
The value $M_{\rm S} = 300\,{\rm GeV}$ corresponds to a minimum possible value to have successful leptogenesis from decays
independently of the initial conditions, so that one would conclude that it is not possible to have dark neutrinos as 
DM and  strong thermal leptogenesis in a unified picture. 
In the next section we will see how the introduction of Higgs portal interactions for the source neutrinos 
drastically changes this conclusion. 

\section{Including Higgs portal interactions for $N_{\rm S}$}

Let us now consider the effect of introducing Higgs portal interactions for the source neutrinos so that the 
Lagrangian in Eq.~(\ref{lagrangian2}) becomes
\be\label{lagrangian3}
-{\cal L}^\n_{Y+M+\L} = \overline{L_{\a}}\,h_{\a J}\, N_{J}\, \widetilde{\Phi} +
                          \frac{1}{2} \, \overline{N^{c}_{I}} \, M_I \, N_{I}  +  
                          {1 \over \widetilde{\L}_{\rm DS}} \, \Phi^\dagger \, \Phi \, \overline{N_{\rm D}^c} \, N_{{\rm S}} + 
                          {1 \over \widetilde{\L}_{\rm SS}} \, \Phi^\dagger \, \Phi \, \overline{N_{\rm S}^c} \, N_{{\rm S}} 
                          + \mbox{\rm h.c.}  \,  ,
\ee
where we introduced the effective scale for the Higgs portal interactions  
$\widetilde{\L}_{\rm SS} \equiv \Lambda/\lambda_{\rm SS}$.

At  zero temperature, after electroweak spontaneous symmetry breaking, they will  yield a contribution 
$\d M_{\rm S}^{\Lambda} = (v^2/\widetilde{\L}_{\rm SS})$ to the source neutrino mass that, however, 
can be safely neglected, anticipating that we will obtain $\widetilde{\L}_{\rm SS} \gtrsim 10^8\,{\rm GeV}$.
At finite temperatures they will give a contribution to the self-energy (see panel (c) in Fig.~1)
and, therefore, to the effective potential, given by
\be\label{yukawapot}
V^{\Lambda}_{\rm SS} =  \frac{T^2}{12\,\widetilde{\Lambda}_{\rm SS}} \,  \;  .
\ee 
This is in addition to the term from the Yukawa couplings in Eq.~(\ref{yukawapot}). In this way the effective Hamiltonian
in Eq.~(\ref{effectiveham}) would now become
\be
\Delta {\cal H} \simeq   
\left( \begin{array}{cc}
- \frac{\D M^2}{4 \, p} - \frac{T^2}{16\,p} \, h^2_{\rm S} -  \frac{T^2}{24\,\widetilde{\Lambda}_{\rm SS}} &  \frac{T^2}{12\,\widetilde{\L}_{\rm DS}}  \\[1ex]
\frac{T^2}{12\,\widetilde{\L}_{\rm DS}} &  \frac{\D M^2}{4 \, p} + \frac{T^2}{16 \, p} \, h^2_{\rm S}  + \frac{T^2}{24\,\widetilde{\Lambda}_{\rm SS}} 
\end{array}\right)   \,  . 
\ee 
This additional term can be comparable or even larger than the term from the Yukawa couplings, depending whether
$T/\widetilde{\Lambda}_{\rm SS} \sim h^2_{\rm S}$ or higher. However, in the relevant range of temperatures where 
dark neutrinos are produced the effect should be limited, at least for large values of $\widetilde{\Lambda}_{\rm SS} \gg 10^{10}\,{\rm GeV}$.  
This is because the dark neutrino production occurs at temperatures much below the resonance \cite{densitym}.  
This conclusion is not changed by  this additional term and, therefore, like also the effective potential from Yukawa couplings, one can expect that
it should not considerably affect the final abundance of dark neutrinos. In any case it represents a subdominant effect compared 
to the contribution of Higgs scatterings to the production of source neutrinos, the main focus of this paper.
For this reason we will neglect it here, though it will be worth to explore its impact in a future paper.

The most important effect of Higgs portal interactions is their contribution to the source neutrino production from 
Higgs scatterings $\phi \, \phi^\dagger \ra N_{\rm S} \, N_{\rm S}$ (diagrammatically shown in the panel (a) of Fig.~3).  
\begin{figure}[t]
\centerline{\psfig{file=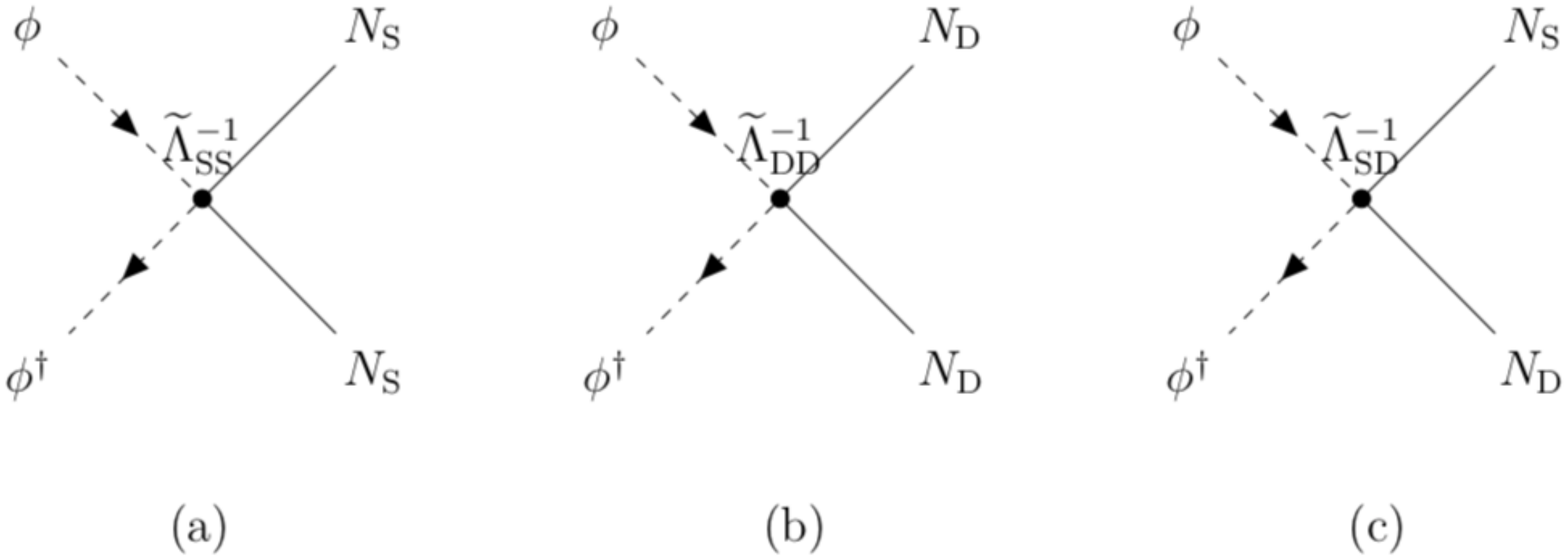,height=15cm,width=11cm,angle=-90}}\vspace{-30mm}
\caption{Higgs portal interactions: source neutrino production (a), dark neutrino production (b) and mixed production (c).}
\end{figure}
In this way Eq.~(\ref{production}) gets generalised into
\be \label{production2}
{d  P_0 \over d z} \simeq {dN_{N_{\rm S}} \over dz} =  - (D+S)\,(N_{N_{\rm S}}-N_{N_{\rm S}}^{\rm eq}) - A \, (N^2_{N_{\rm S}} - N^{{\rm eq}\,2}_{N_{\rm S}}) \,   ,
\ee
where we defined
\be
A \equiv {\langle \sigma_{\phi\phi^\dagger \ra N_{\rm S}N_{\rm S}^c}\, v_{\rm rel} \rangle \over H(z)\,z \, R^3(z)} \,  ,
\ee
with $\langle \sigma_{\phi\phi^\dagger \ra N_{\rm S}N_{\rm S}^c}\, v_{\rm rel} \rangle$ the thermal averaged cross section.
Here $R^3(z)$ is the portion of comoving volume where abundances are calculated, essentially a normalisation factor. 
With our choice,  $N_{N_{\rm S}}^{\rm eq}(z_{\rm S} \ll 1) = 1$, one has
\be
R^3(z) = {4\over 3}\,{\pi^2 \over g_{N_{\rm S}}\zeta(3)}\,{z^3 \over M_{\rm D}^3} \,   .
\ee   
In the regime $M_{\rm S} \ll T$ we are interested in, the thermal averaged cross section is simply given by \cite{Kolb:2017jvz}
\be
\left.\langle \sigma_{\phi\phi^\dagger \ra N_{\rm S}N_{\rm S}}\,v_{\rm rel} \rangle\right|_{M_{\rm S}\ll T} 
= {1 \over 4 \pi \, \widetilde{\Lambda}_{\rm SS}^{2} } \,  .
\ee
Combining all pieces together, we obtain:
\be
A(z) = {A_1 \over z^2} \,  , \;\; \mbox{\rm with} \;\;
A_1 \equiv A(z=1) = {3 \over 16}\, {\zeta(3) \over \pi^3}\,\,g_{N_{\rm S}}\,\sqrt{90\over 8\,\pi^3\,g_R}\, 
{M_{\rm D}\, M_{\rm P}\over \widetilde{\Lambda}_{\rm SS}^{2}} \,  .
\ee
It is also convenient to give a numerical expression for $A_1$,
\be\label{A1num}
A_1 \simeq 1.0 \times 10^{-11}\,\left({M_{\rm D}\over 100\,{\rm TeV}}\right)\,
\left({10^{16}\,{\rm GeV}\over \widetilde{\Lambda}_{\rm SS}}\right)^2 \,  ,
\ee
where we used the SM value $g_R = 106.75$ for the number of ultrarelativistic degrees of freedom and $g_{N_{\rm S}}=2$. 
It is simple to find an approximate solution for $N_{N_{\rm S}}(z)$, valid in the regime $N_{N_{\rm S}} \ll N_{N_{\rm S}}^{\rm eq}$
and for $z \ll 1$, so that we can take $N_{N_{\rm S}}^{\rm eq} \simeq 1$. 
Assuming that the production from Higgs portal interactions dominates and taking 
as initial condition $N_{N_{\rm S}}(z_{\rm in}) = 0$, one immediately finds
\be
N_{N_{\rm S}}(z) \simeq A_1 \, \left({1 \over z_{\rm in}} - {1 \over z}\right) \,  ,
\ee
implying asymptotically, but still for $z \ll 1$, 
\be\label{NSabundance}
N_{N_{\rm S}}(z_{\rm in} \ll z \ll 1) \simeq {A_1\over z_{\rm in}} \simeq 1.0 \times \left({T_{\rm in}\over 10^{16}\,{\rm GeV}}\right) 
\,  \left({10^{16}\,{\rm GeV} \over \widetilde{\Lambda}_{\rm SS}}\right)^2 \,  ,
\ee
where the numerical expression has been obtained from Eq.~(\ref{A1num}) and where notice that the dependence on $M_{\rm D}$
has cancelled out.
This expression of course is valid only for $N_{N_{\rm S}}(z_{\rm in} \ll z \ll 1) \ll 1$, so that it can be also read as 
a condition on $T_{\rm in}$ and $\widetilde{\Lambda}_{\rm SS}$ for the thermalisation of the source neutrinos prior to the onset
of source-dark neutrino oscillations. In this case
it can also be recast simply as $\widetilde{\Lambda}_{\rm SS}^2/T_{\rm in} =10^{16}\,{\rm GeV}$.  
Notice that, for the validity of the effective theory, one has to impose $T_{\rm in} \lesssim \widetilde{\Lambda}_{\rm SS}$.

As initial temperature we can assume $T_{\rm in} = T_{\rm RH}$, where $T_{\rm RH}$ is the reheat temperature below which one
can assume a radiation dominated regime.\footnote{We do not include a production between $T_{\rm max} > T_{\rm RH}$ and $T_{\rm RH}$
\cite{Kolb:2017jvz} since this would just very slightly contribute to relax the final constraints.}
The result we obtained shows that  for $\widetilde{\Lambda}_{\rm SS} \leq 10^{16}\,{\rm GeV}$ 
one can always obtain a full thermalisation of the source neutrino abundance by  increasing $T_{\rm RH}$. 
In particular, in the limit case $\widetilde{\Lambda}_{\rm SS} = 10^{16}\,{\rm GeV}$,
the thermalisation can be obtained for the maximum value allowed by cosmological observations $T_{\rm RH} =10^{16}\,{\rm GeV}$.

In Fig.~4 we show, for the same values of $M_{\rm S}$, $M_{\rm D}$ and $\tau_{\rm D}$ as in Fig.~2, 
the evolution of $N_{N_{\rm S}}$ and $N_{N_{\rm D}}$ when Higgs portal interactions for $N_{\rm S}$  are taken into account. 
In the upper panel we fixed 
$\widetilde{\Lambda}_{\rm SS} =10^{16}\,{\rm GeV}$ and we show solutions for the indicated  values of $T_{\rm RH}$. 
\begin{figure}[t]
\centerline{\psfig{file=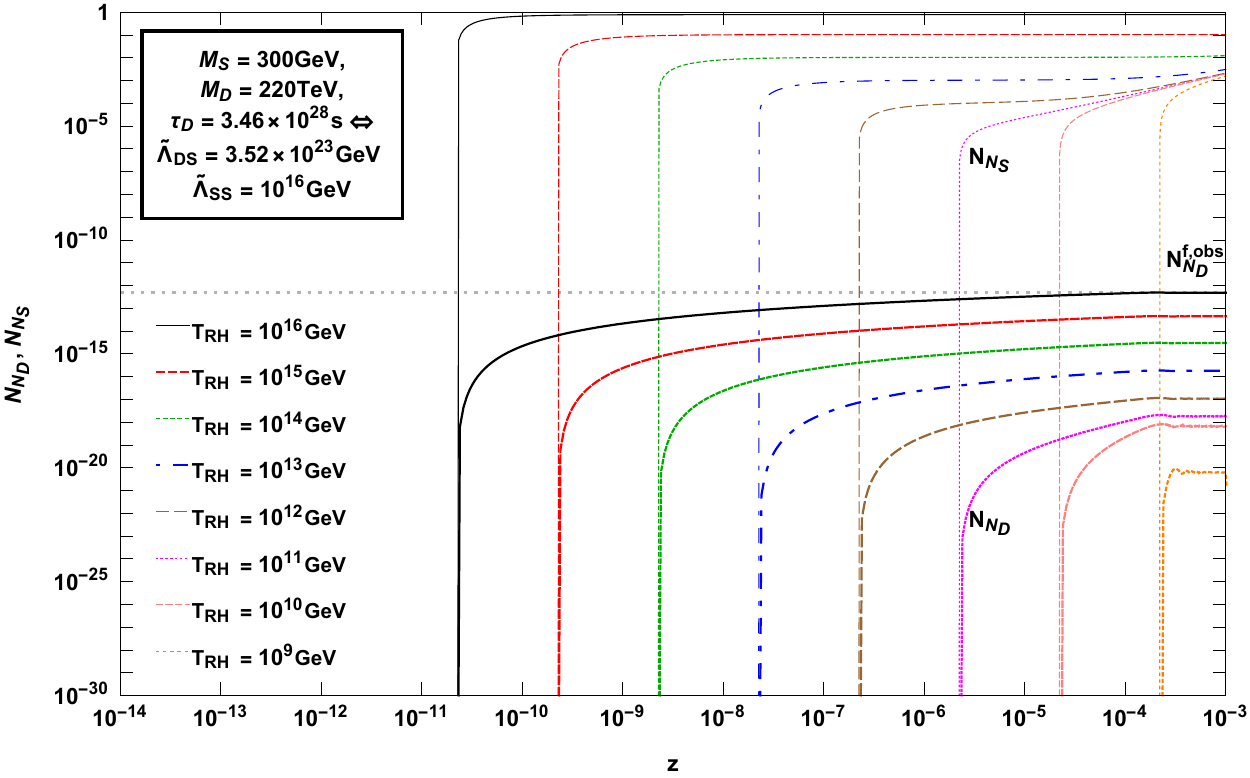,height=6cm,width=9cm,angle=0}}
\vspace{5mm}
\centerline{\psfig{file=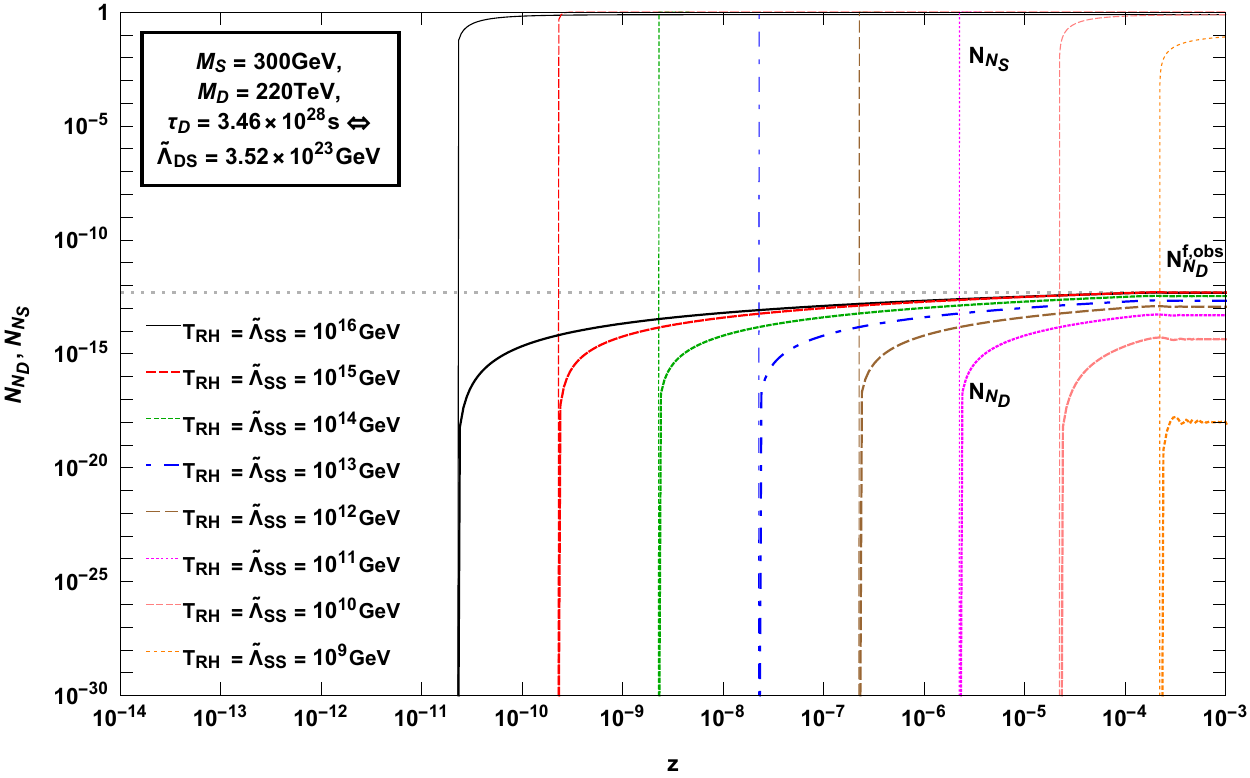,height=6cm,width=9cm,angle=0}}
\caption{Effect of source neutrino Higgs portal interactions on the source and dark neutrino abundances.}
\end{figure}
One can clearly see how the source neutrino abundance thermalises for the maximum value 
$T_{\rm RH} = 10^{16}\,{\rm GeV}$, while it is suppressed linearly for lower values. Of course it should be noticed how,
for $T_{\rm RH} = 10^{16}\,{\rm GeV}$, one obtains a DM abundance that is greater than the observed value, clearly indicating that
one can always find choices of the parameters reproducing the observed value. 

In the lower panel  we have conservatively imposed $T_{\rm RH} = \widetilde{\Lambda}_{\rm SS}$ and we show
the evolution of the source and dark neutrino abundances for the indicated values of
$\widetilde{\Lambda}_{\rm SS} \leq  10^{16}\,{\rm GeV}$. Notice how in this case the thermalisation condition is always respected.
Therefore, one obtains solutions that are clearly very close to those shown in the upper panel of Fig.~2 with the difference
that now the thermalisation is not just assumed but obtained as the result of the Higgs portal interactions for the source neutrino. 

Finally, let us highlight again that we are neglecting a possible production, pure or mixed, of dark neutrinos directly from  
Higgs portal interactions (shown diagrammatically in the panel (b) and panel (c) of Fig.~3).  This 
is equivalent to say that their corresponding effective scales $\widetilde{\L}_{\rm DD} \equiv \Lambda/\lambda_{\rm DD}$ 
and $\widetilde{\L}_{\rm DS}$,  are sufficiently large that the associated production is negligible compared to the contribution from neutrino mixing.  
It is simple  to derive  a lower bound on $\widetilde{\L}_{\rm DD}$ and $\widetilde{\L}_{\rm DS}$ 
imposing that the dark neutrino relic abundance produced from Higgs portal interactions 
is negligible compared to the  observed abundance in Eq.~(\ref{NDobs}). This abundance can be calculated in the same way as the source neutrino
abundance and therefore it will be given by Eq.~(\ref{NSabundance}) simply replacing $\widetilde{\Lambda}_{\rm SS}$ 
with  $\widetilde{\Lambda}_{\rm DD}$ or $\widetilde{\Lambda}_{\rm DS}$. One can then easily derive the conditions
\be\label{LDDcond}
\widetilde{\Lambda}_{\rm DD},  \widetilde{\Lambda}_{\rm DS}\gg 10^{22}\,{\rm GeV} \, 
\sqrt{ {T_{\rm RH}\over 10^{16}\,{\rm GeV}}\, {M_{\rm D}\over {\rm PeV}} } \,  .
\ee
  We will discuss in Section 5 a model where these two conditions are naturally satisfied.

\section{Allowed regions in the dark neutrino lifetime-mass plane and experimental constraints}
   
 We determined the allowed regions in the dark neutrino lifetime-mass plane and we show the results in Figs.
 5--8. In the first case, as in Fig. 1, we fix $\widetilde{\Lambda}_{\rm SS} = 10^{\rm 16}\,{\rm GeV}$
 and show the allowed regions for different values of the reheat temperature, with $T_{\rm RH} \leq \widetilde{\Lambda}_{\rm SS}$, as indicated.
 The different panels are for $M_{\rm S} = 300\,{\rm GeV}$ (top) and $1\,{\rm TeV}$ (bottom) in Fig.~5
and for  $M_{\rm S} = 10\,{\rm TeV}$ (top) and $100\,{\rm TeV}$ (bottom) in Fig.~6.
 The allowed regions are obtained imposing $N_{N_{\rm D}}\geq N_{\rm D}^{\rm f,obs}$. On the borders one has exactly
 $N_{N_{\rm D}} = N_{\rm D}^{\rm f,obs}$, while any point inside would corresponds to an overabundance 
 but this can be lowered simply lowering $T_{\rm RH}$ and/or increasing $M_{\rm S}$. 
 
 In each panel the  shadowed region for $\tau_{\rm D} \leq 10^{28}\,{\rm s}$ is indicatively the region that is currently tested at 
 neutrino telescopes, such as IceCube, and gamma ray telescopes, such as the Fermi Gamma-Ray telescope, placing
 lower bounds on the lifetime. An accurate lower bound in fact depends on the mass of the DM particle
 and on the specific primary decay channel. It also depends on a description of the astrophysical contribution that in this case
 plays the role of a background. A positive signal should show up as an excess with respect to this astrophysical background. 
 We show some of the lower bounds on the lifetime of a decaying DM recently obtained by the
 IceCube collaboration at $90\%\,{\rm C.L.}$ \cite{IceCube:2022vtr}. 
 The thin solid line indicates the lower bound on the DM mass in the range $160\,{\rm TeV}$--$20\,{\rm PeV}$ 
 obtained from 7.5 yr High-Energy Starting Event (HESE) data  in the energy range 60 TeV to 10 PeV 
 for the decay channel ${\rm DM} \ra {\rm Higgs} + \nu$. 
 We also show, with a thick solid line, the same lower bound but for the decay channel ${\rm DM} \ra b + \bar{b}$. 
 At lower masses we also indicate, with a dashed line, a lower bound obtained from 2 yr cascade events  
 for the decay channel ${\rm DM} \ra \mu\,\bar{\mu}$  and at even lower masses, with a dotted line, the lower bound 
 from the Fermi gamma-ray telescope for the channel ${\rm DM} \ra \nu \, \bar{\nu}$ also reported in \cite{IceCube:2022vtr}. 
 We also show other recent lower bounds obtained from gamma ray observations in different decay channels. With a thick dotted line
 we show the lower bound obtained in \cite{Chianese:2021jke} 
 for the decay channel ${\rm DM} \ra b + \bar{b}$  combining data from ultra-high-energy gamma-ray measurements
 and with a thin dotted line the lower bound obtained in \cite{LHAASO:2022gsi} 
  analysing data from the Large High Altitude  Air Shower Observatory (LHAASO). 
 Of course our case is different from the considered decay channels since the
 dark neutrino can decay both into Higgs boson and gauge bosons plus neutrino or charged leptons.   
All shown lower bounds should then be regarded as indicative and a dedicated analysis would be needed. 
\begin{figure}[t]
\centerline{\psfig{file=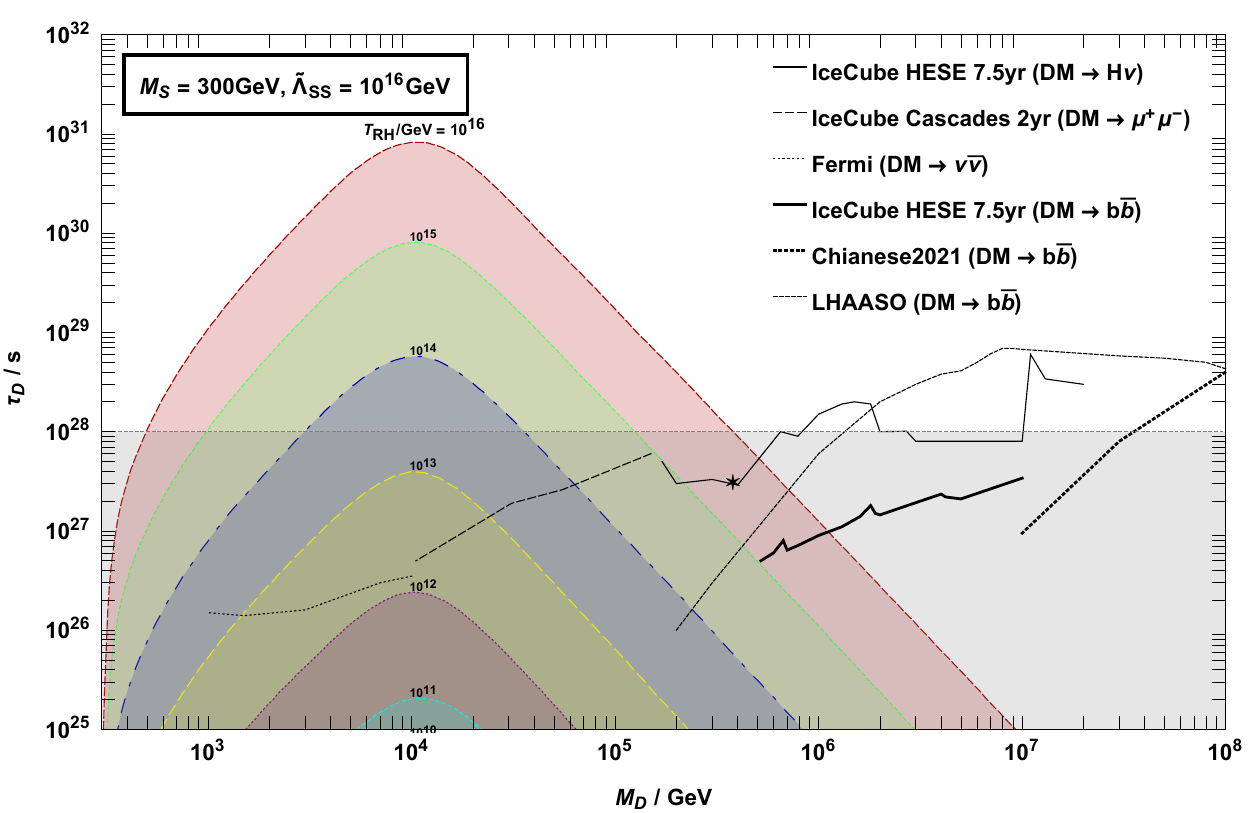,height=6cm,width=10cm,angle=0}}
\vspace{4mm}
\centerline{\psfig{file=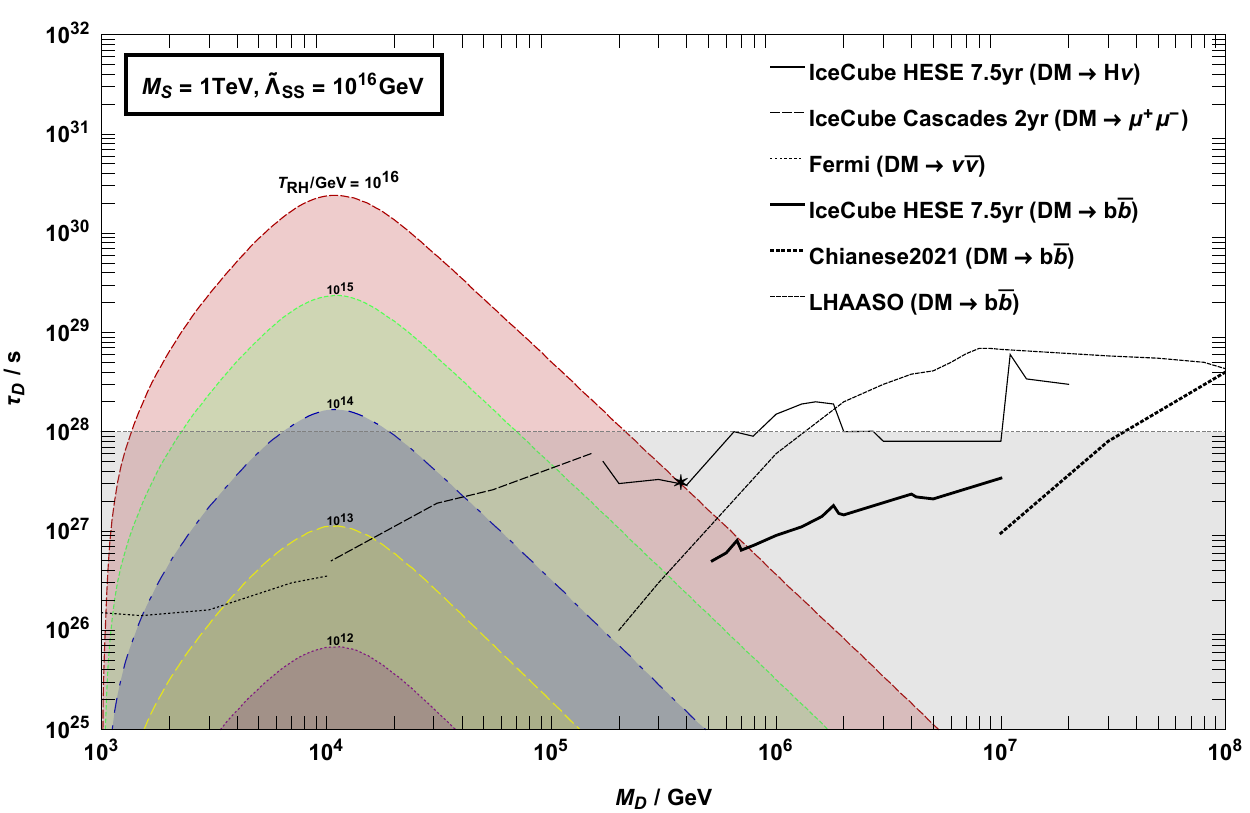,height=6cm,width=10cm,angle=0}}
\caption{Allowed regions in the lifetime versus mass plane for a fixed value $\tilde{\Lambda}_{\rm SS} = 10^{16}\,{\rm GeV}$, 
for the indicated values of $T_{\rm RH}$ and for $M_{\rm S}=300\,{\rm GeV}$ (upper panel), $1\,{\rm TeV}$ (bottom panel).}
\end{figure}
  Interestingly, the  IceCube collaboration also confirms the presence of an excess at ${\cal O}(100\,{\rm TeV})$ energies compared to an
 astrophysical component. The decaying DM hypothesis improves the data fit with a 2.5$\s$ statistical significance and the best fit
 is found for the decay channel ${\rm DM} \ra b \bar{b}$ with $M_{\rm DM} =289\,{\rm TeV}$ and $\tau_{\rm DM}=2.8 \times 10^{27}\,{\rm s}$. 
 This best fit point is denoted by a star in the panels and it should also be regarded as indicative. It would be of course interesting to see whether
 a dedicated analysis within the RHINO model can also address the excess and the statistical significance of the solution.     
\begin{figure}[t]
\centerline{\psfig{file=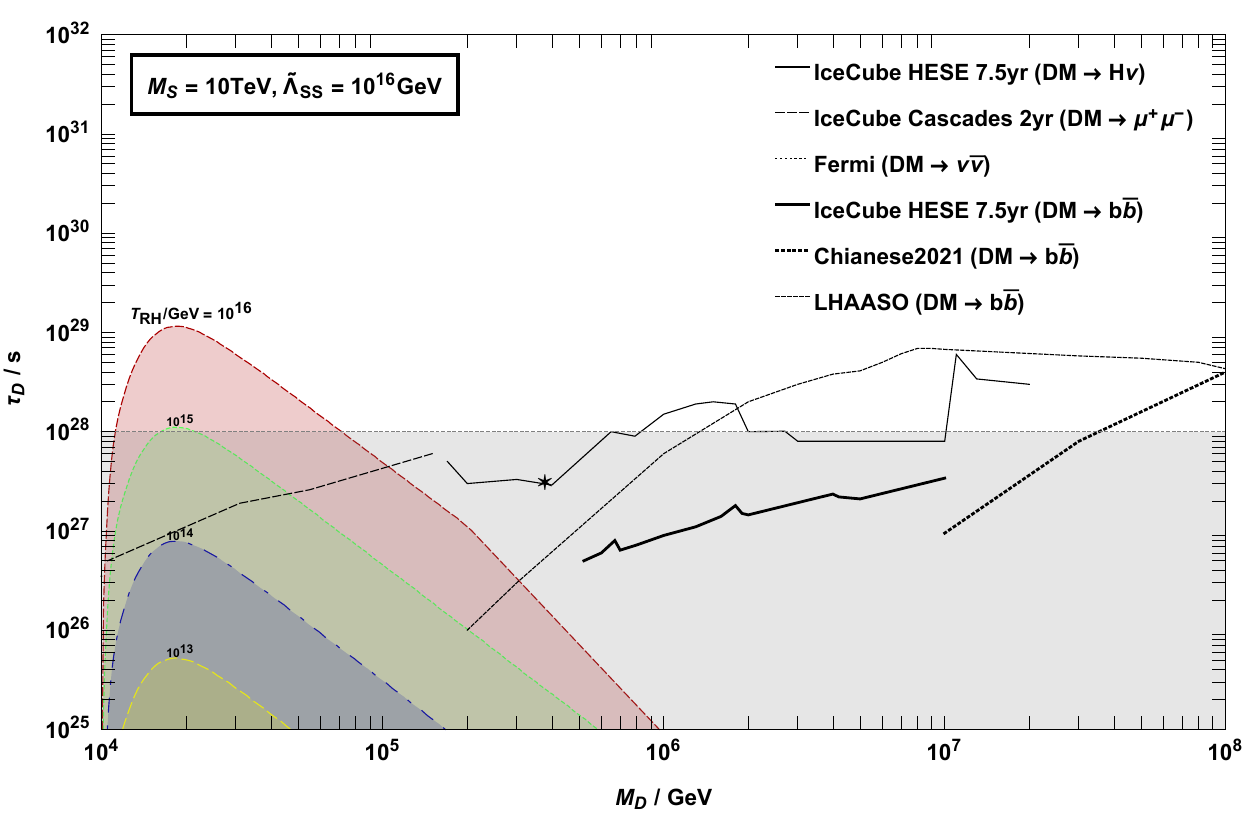,height=6cm,width=10cm,angle=0}}
\vspace{4mm}
\centerline{\psfig{file=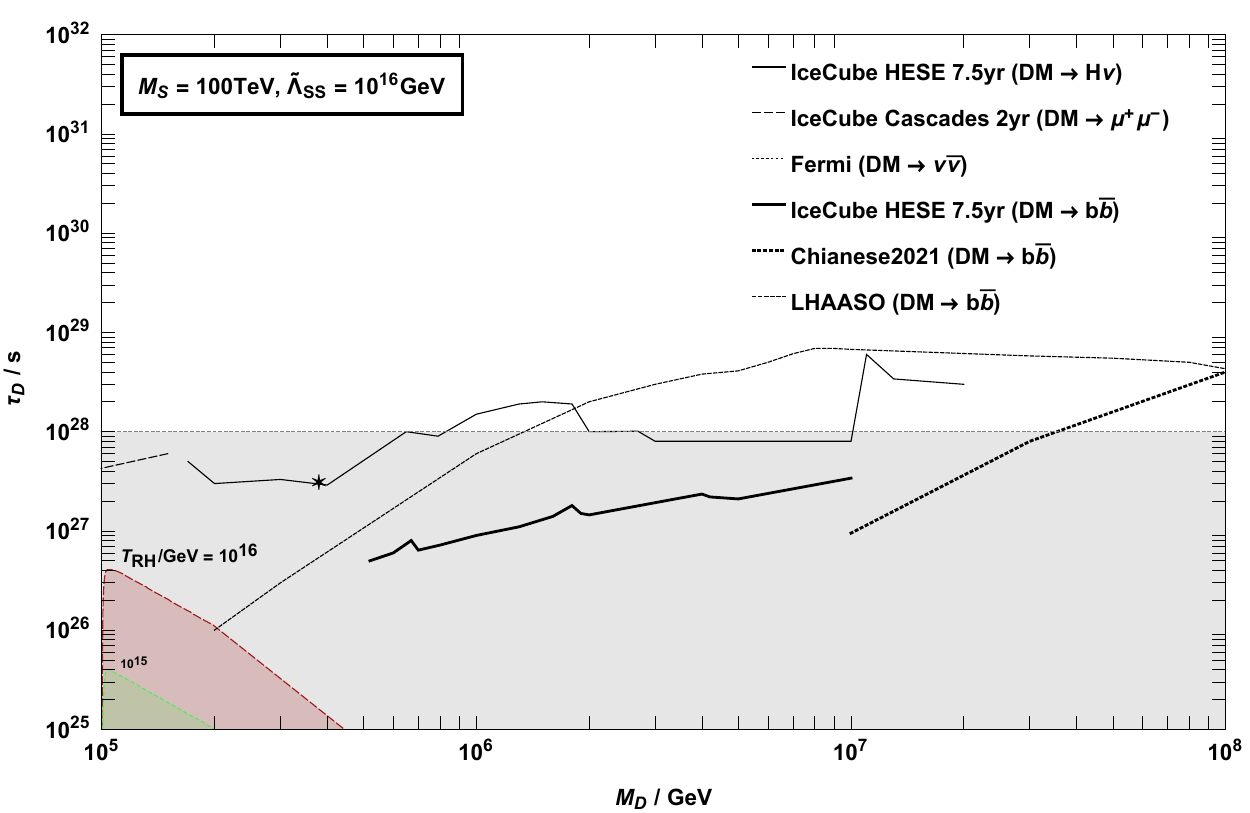,height=6cm,width=10cm,angle=0}}
\caption{Allowed regions in the lifetime versus mass plane for a fixed value $\tilde{\Lambda}_{\rm SS} = 10^{16}\,{\rm GeV}$, 
for the indicated values of $T_{\rm RH}$ and for $M_{\rm S}=10\,{\rm TeV}$ (upper panel), $100\,{\rm TeV}$ (bottom panel).}
\end{figure}

 In Figs.~7 and 8 we show the allowed regions, respectively, 
 for the indicated values of $\widetilde{\Lambda}_{\rm SS} = T_{\rm RH}$ and again 
 for $M_{\rm S} = 300\,{\rm GeV}$ (top) and $1\,{\rm TeV}$ (bottom) in Fig.~7
and for  $M_{\rm S} = 10\,{\rm TeV}$ (top) and $100\,{\rm TeV}$ (bottom) in Fig.~8.

\begin{figure}[t]
\centerline{\psfig{file=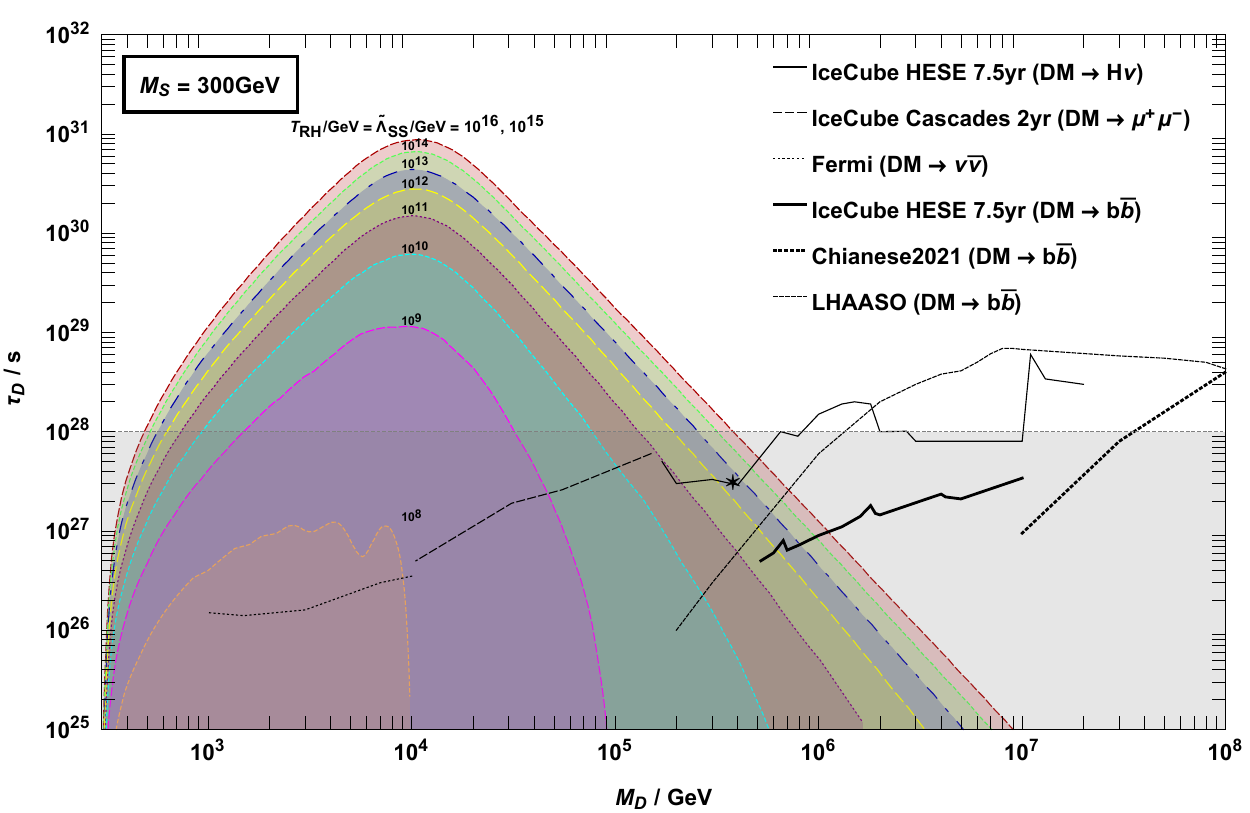,height=6cm,width=10cm,angle=0}}
\vspace{4mm}
\centerline{\psfig{file=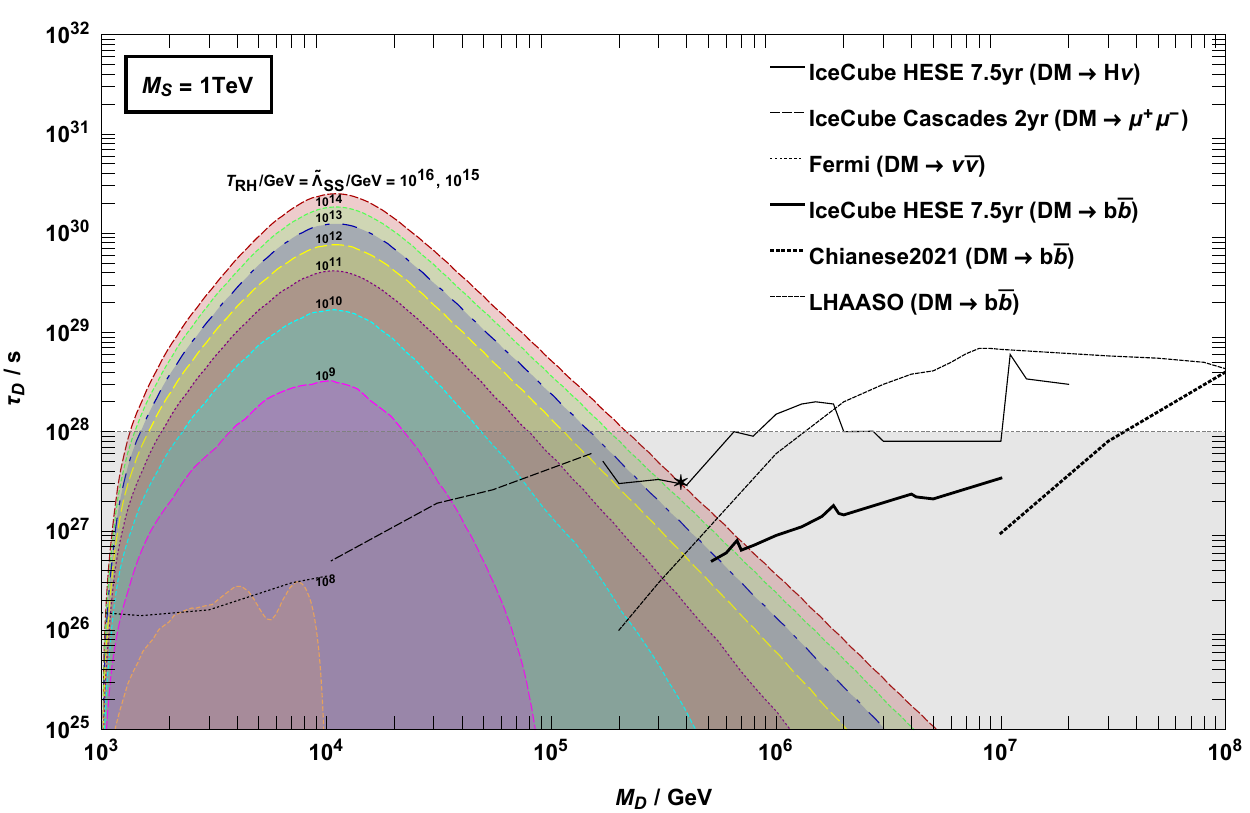,height=6cm,width=10cm,angle=0}}
\vspace{2mm}
\caption{Allowed regions in the lifetime versus mass plane for the indicated values of
$\tilde{\Lambda}_{\rm SS} = T_{\rm RH}$ for $M_{\rm S} = 300\,{\rm GeV}$ (top) 
and $M_{\rm S} = 1\,{\rm TeV}$.}
\end{figure}

\begin{figure}[t]
\centerline{\psfig{file=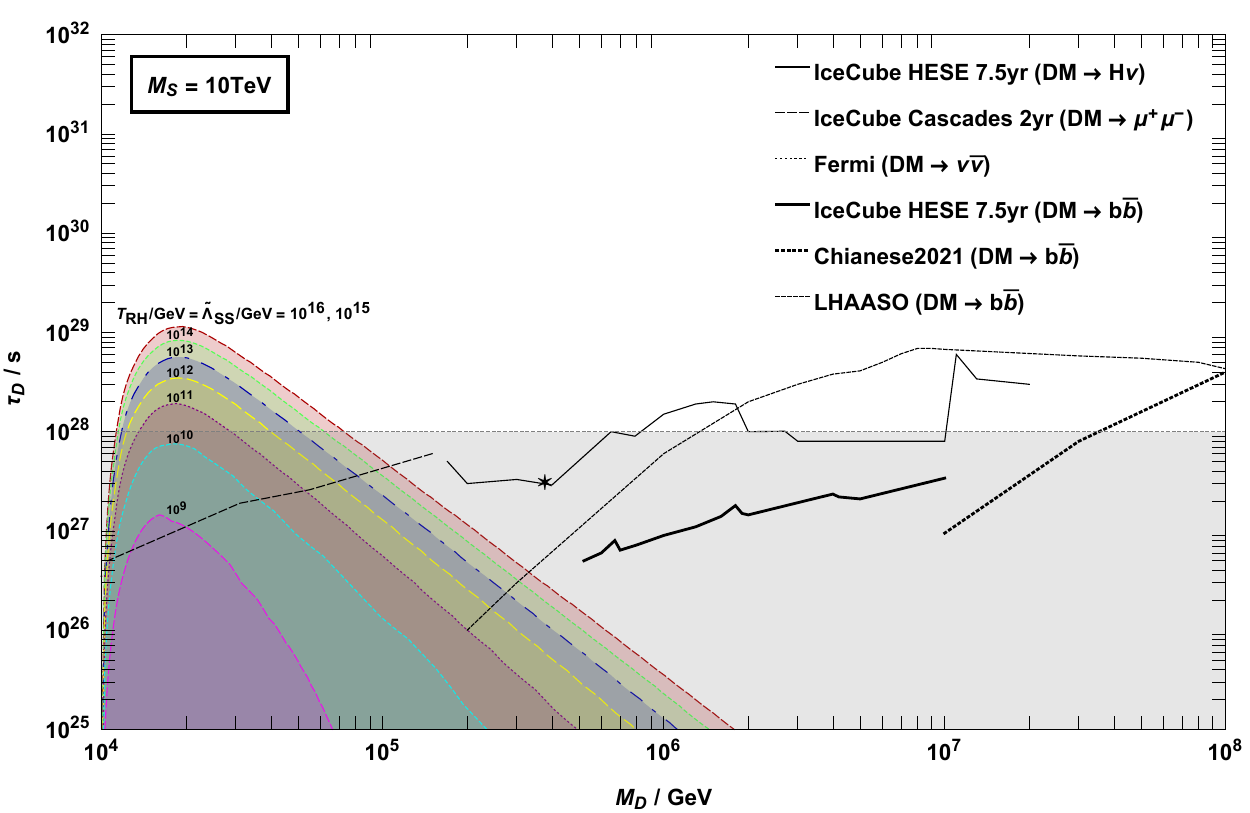,height=6cm,width=10cm,angle=0}}
\vspace{4mm}
\centerline{\psfig{file=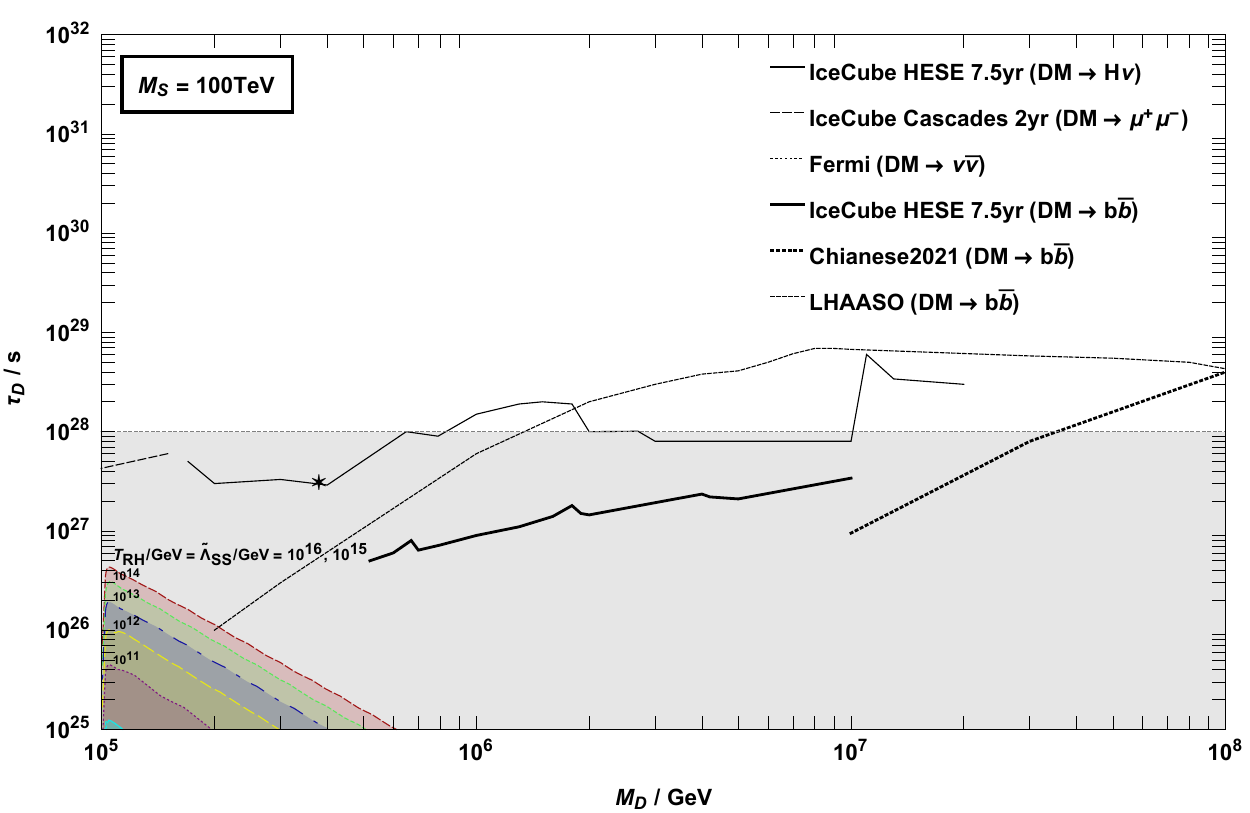,height=6cm,width=10cm,angle=0}}
\caption{As in Fig.~7 but for $M_{\rm S} = 10\,{\rm TeV}$ (top) and $M_{\rm S} = 100\,{\rm TeV}$ (bottom).}
\end{figure}
One can see that, as far as $\widetilde{\Lambda}_{\rm SS} = T_{\rm RH} \gg 10^{10}\,{\rm GeV}$, the allowed regions get
only slightly reduced when $\widetilde{\Lambda}_{\rm SS}$ decreases. On the other hand, for  $\widetilde{\Lambda}_{\rm SS} = T_{\rm RH} \lesssim 10^{10}\,{\rm GeV}$, the  reduction becomes quite significant and for  $\widetilde{\Lambda}_{\rm SS} = T_{\rm RH} \lesssim 10^{8}\,{\rm GeV}$ 
there is basically no allowed region, showing that $T_{\rm RH} \gtrsim 10^8 \, {\rm GeV}$ can be regarded as a conservative lower bound 
for $T_{\rm RH}$, of course under the assumption that dark neutrinos play the role of DM. 
This result is a consequence of what we have already noticed discussing the dark neutrino abundance evolution shown in
the bottom panel of Fig.~4: since the onset of neutrino oscillations, and consequently the dark neutrino production,  
occurs approximately at $T \sim 10^{9}\,{\rm GeV}$, as far as the reheat temperature is much greater than this value 
there is no much variation when $T_{\rm RH}$ is increased, but for lower values the final dark neutrino abundance gets
strongly suppressed.  Notice that such a lower bound on the reheat temperature would be compatible with the well known 
upper bound $T_{\rm RH} \lesssim 10^{10}\,{\rm GeV}$ from the gravitino problem in gravity mediated supersymmetric models \cite{gravitino}. 
The same lower bound also holds on the effective scale $\widetilde{\L}_{\rm SS} \gtrsim 10^8 \, {\rm GeV}$. 
This lower bound confirms the validity of having neglected the Majorana mass term originating from 
the Higgs portal operator for  the source neutrino in Eq.~(\ref{angen}), since one has
$|\d M_{\rm S}^{\Lambda}| \lesssim v^2/\widetilde{\Lambda}_{\rm SS} \lesssim  0.1 \,{\rm MeV}$. 
 
It should also be noticed that for  $\widetilde{\Lambda}_{\rm SS} = T_{\rm RH} = 10^{16}\,{\rm GeV}$, 
the value that maximises the dark neutrino final abundance, there are allowed regions only for $M_{\rm S} \lesssim 100\,{\rm TeV}$.
This should be regarded as an upper bound of the model on the seesaw scale.\footnote{However, notice that our analysis 
assumes $M_{\rm D} \geq M_{\rm S}$. It should be understood whether there can be solutions also 
in the case $M_{\rm S} > M_{\rm DM}$, this interesting possibility requires a dedicated study and 
will be explored elsewhere.} 
 
\section{UV-completing RHINO}         
 
Let us now finally discuss two possible  UV-complete RHINO models that were already qualitatively sketched in the conclusions of \cite{ad}.
In the first case, the mediator in the Anisimov operators in Eq.~(\ref{anisimov}) is a heavy scalar $H$
with vanishing vev. In the second case, the mediator is a heavy fermion $F$.\footnote{Of course there
could be more than one heavy fermion, the generalisation is straightforward.}

\subsection{Heavy scalar $H$ as mediator}

Let us consider an extension of the seesaw Lagrangian where a heavy real scalar field $H$ (with vanishing vev) is introduced and couples to
the RH neutrinos with Yukawa couplings $y_{IJ}$ and to the standard Higgs field with a trilinear coupling $\mu$:\footnote{This model
was also discussed in \cite{unified,Kolb:2017jvz}.}
\be
{\cal L}_H = {1\over 2}\partial_\mu H \partial^\mu H -{1\over 2}\,M^2_H \, H^2 - \sum_{I,J} \, \lambda_{IJ}\,H \, \overline{N_{\rm I}^c} \, N_{J} 
- \mu \, H \, \phi^\dagger \, \phi \,  .
\ee
At scales much below $M_H$ we can integrate out $H$, obtaining the effective Lagrangian
\be
{\cal L}_H^{\rm eff}= 
{1\over 2}\,\sum_{I,J,K,L} {\lambda_{IJ}\lambda_{KL}\over M^2_H} \, (\overline{N_{\rm I}^c} \, N_{J})\,(\overline{N_{\rm K}^c} \, N_{L}) 
+ {1\over 2}\,{\mu^2 \over M^2_H}\,(\phi^\dagger \, \phi)^2
+ \sum_{I,J} {\mu\,\lambda_{IJ} \over M^2_H}\, \, \Phi^\dagger \, \Phi \, \overline{N_{\rm I}^c} \, N_{\rm J} \,  .
\ee
One can clearly recognise the Anisimov operators in Eq.~(\ref{angenprime}) where the effective scales can 
be identified with $\widetilde{\Lambda}_{IJ} = \Lambda/\lambda_{IJ}$,
and $\Lambda = M^2_H/\mu$.  Diagrammatically, the self-energy diagram in the panel (b) of  Fig.~1
and the scattering diagram in panel (a) of Fig.~3 are obtained by the diagrams in Fig.~9, panel (a) and panel (b),  respectively. 
\begin{figure}[t]
\centerline{\psfig{file=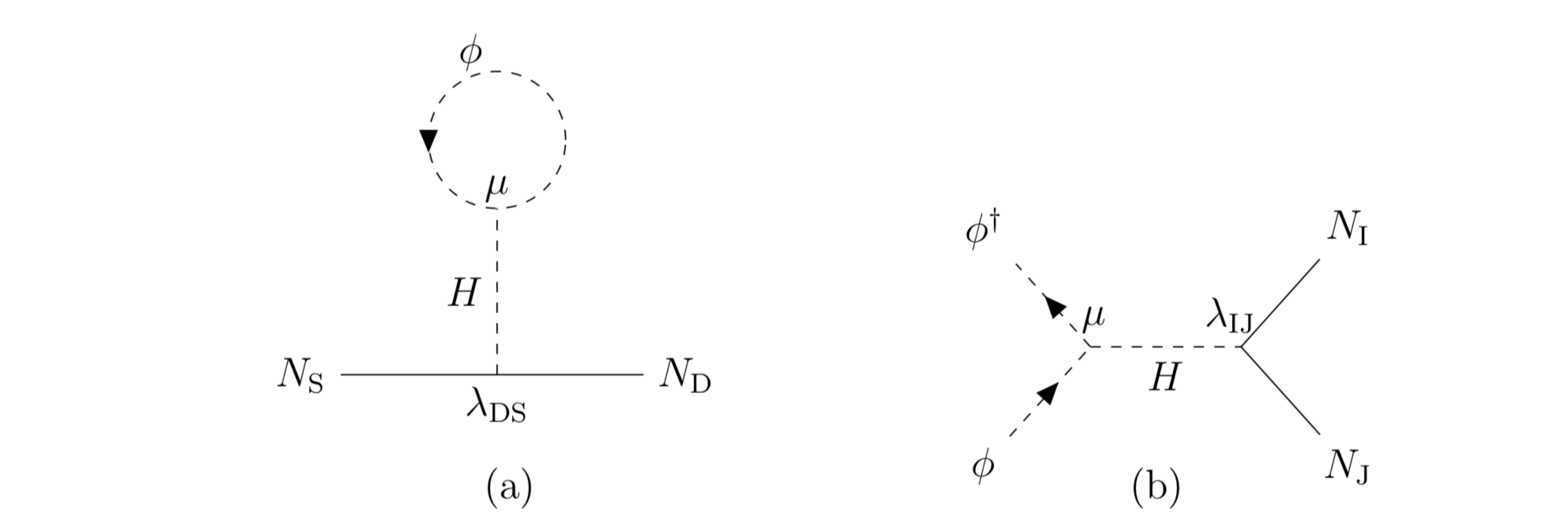,height=6.5cm,width=19cm,angle=0}}\vspace{-2mm}
\caption{Feynman diagrams with a heavy scalar $H$ as mediator and $I,J={\rm D,S}$. Integrating out $H$, they 
lead to the Feynman diagrams in panel (b) of Fig.~2 and  (a), (b) and (c) of Fig.~3.}
\end{figure}
The appealing feature of this model is that one can get a trans-Planckian value for the effective scale $\widetilde{\L}_{\rm DS} \sim 10^{23}\,{\rm GeV}$ 
even for $\lambda_{IJ} = {\cal O}(1)$, simply choosing $\mu \ll M_{\rm GUT}$, for example, $M_H \sim M_{\rm GUT} \sim 10^{16}\,{\rm GeV}$
and $\mu \sim 10^9\,{\rm GeV}$. However, the problem of this setup is that in this case one cannot also reproduce the effective scale 
$\widetilde{\Lambda}_{\rm SS} \sim 10^{16}\,{\rm GeV}$ for the source neutrino Higgs portal interactions. 
In  that respect, one should  arbitrarily assume 
$\Lambda \sim 10^{16}\,{\rm GeV}$, for example for $\mu = M_H \sim M_{\rm GUT} \sim 10^{16}\,{\rm GeV}$, 
 $\lambda_{\rm DS} \sim 10^{-7}$ and $\lambda_{\rm SS}\ll 10^{-7}$ in order for  
$\widetilde{\Lambda}_{\rm DD}$ to satisfy the condition in Eq.~(\ref{LDDcond}).
 
However, there is a much simpler model where one can nicely understand the values of the effective scales that, as we have seen, would be able
to address the DM problem compatibly with successful strong thermal leptogenesis and experimental constraints from neutrino telescopes. 

\subsection{Heavy fermion $F$ as mediator}

 Let us this time extend the seesaw Lagrangian introducing an heavy fermion doublet $F$ with Yukawa couplings $y_I$ to RH neutrinos, 
 explicitly:
 \be
 {\cal L}_F = \bar{F}\, ( i\, \slashed{\partial} - M_{\rm F} ) \, F - 
 \sum_I \, y_{I}\, (\bar{F} \, \phi \, N_{I} + \bar{N}_I\,\phi^\dagger \, F) \,  .
 \ee
At scales much below $M_{\rm F}$ one can integrate out  $F$ obtaining the effective Lagrangian
\be
-{\cal L}_{F}^{\rm eff} = \sum_{I,J} {y_I \, y_J \over M_F} \, \bar{N}_I\,N_J\,\phi^\dagger\,\phi \,  ,
\ee
where the RH side coincides with the Anisimov operators with the simple identification $\L = M_{\rm F}$
and $\lambda'_{IJ} = y_I\,y_J$. The three Anisimov operators in Eq.~(\ref{angen}), Higgs-induced neutrino mixing, source neutrino
Higgs portal interactions and dark neutrino Higgs portal interactions, can then be regarded as the low energy effective operators
generated  by the three diagrams in Fig.~10, respectively.
\begin{figure}[t]
\centerline{\psfig{file=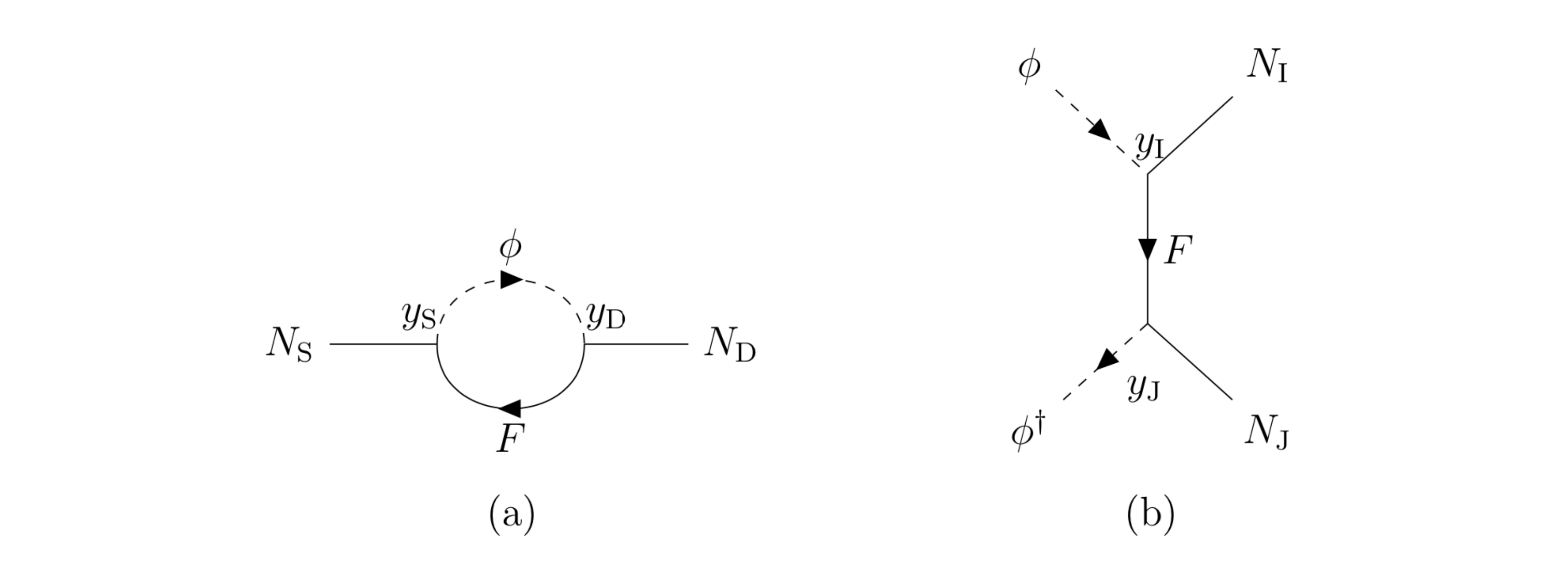,height=7cm,width=19cm,angle=0}}\vspace{0mm}
\caption{Feynman diagrams with a heavy fermion $F$ as mediator and $I,J={\rm D,S}$. Integrating out $F$, they 
lead to the Feynman diagrams in panel (b) of Fig.~2 and in panels (a) and (b) of Fig.~4.}
\end{figure}

If we take $M_{\rm F} \sim M_{\rm GUT}$, $y_{\rm S} \sim 1$ and $y_{\rm D} \sim 10^{-7}$, one can immediately 
reproduce the values $\widetilde{\L}_{\rm SS} \sim 10^{16}\,{\rm GeV}$,  $\widetilde{\L}_{\rm DS} \sim 10^{23}\,{\rm GeV}$ 
and $\widetilde{\L}_{\rm DD} \sim 10^{30}\,{\rm GeV}$. As we have seen, these are the correct values to reproduce the 
observed DM abundance from Higgs-induced RH neutrino mixing, with source neutrino Higgs portal interactions able to thermalise the source neutrino abundance prior to the onset of the oscillations and with a  suppressed contribution to dark neutrino production that we have indeed neglected. 
These values are  much less arbitrary than the choice in the previous model, since the three couplings $\lambda_{IJ}$ are the product of just two Yukawa couplings and it is non trivial that the third is obtained automatically and satisfying correctly the condition  in Eq.~(\ref{LDDcond}). Moreover, 
they can be well understood imposing a ${\mathbb Z}_2$ symmetry under which all particles are even, except
the dark neutrino that is odd. In this way the small Yukawa coupling $y_{\rm D} \sim 10^{-7}$ could be regarded as a small symmetry 
breaking parameter connecting the visible sector to the dark sector.

\section{Conclusions}        

We have seen how, including Higgs portal interactions for the source neutrino, the RHINO model can fully express its potential,
providing a model for a decaying heavy DM particle in the range $1\,{\rm TeV}$--$1\,{\rm PeV}$, 
compatible with strong thermal resonant leptogenesis and testable at neutrino telescopes.
 In this way RHINO can be regarded as quite a minimal model of the origin of matter and neutrino masses.
Since the leptogenesis scale can be higher than the sphaleron freeze-out temperature, the 
final matter-antimatter asymmetry is independent of the initial conditions. Notice, moreover, that the dark neutrino abundance
is independent of a possible external contribution to the production of the source neutrino abundance, 
since this is anyway thermalised by Higgs portal interactions. On the other hand, of course, 
it is not independent of a possible  additional direct production of the dark neutrinos from some external mechanism, that, therefore, has to be 
assumed to be negligible or in any case sub-dominant.
In this respect the main competitive mechanism is a possible gravitational production. However, typically, this is non negligible only for
even heavier particles, for example in the case of WIMPzillas \cite{Kolb:2017jvz,Chung:1998ua}. Therefore,  RHINOs and WIMPzillas seem to
be successful candidates of DM particles in different  mass ranges.  In addition, based just on cosmological considerations, 
we obtained as an extra attractive feature of the model that the natural fundamental scale for the effective interactions,
responsible within RHINO for the production and the decay of the DM,  is the grandunified scale. 
We have also seen how our results point to a simple  UV-complete model where the mediator of the Anisimov interactions is a heavy fermion. 
It should be appreciated how this UV-complete model  simultaneously yields correctly the three effective scales in the Anisimov operators
when the fundamental scale is identified with the grandunified scale (or close to it). 
Our results also show that the RHINO model can nicely address the current  hint from 7.5yr HESE IceCube data for a 100 TeV excess 
in the high energy neutrino flux on the top of an astrophysical component with spectral index $\gamma \simeq 2.2$.  
If an explanation in terms of a decaying DM is correct, then gradually this excess should exhibit anisotropies tracking current
DM distribution, since the signal would be simply proportional to the DM density. We believe that the RHINO model
would be the leading candidate to explain such an excess, since  it is the only model that has genuinely {\em pre}-dicted  such a signal \cite{ad},
since the same physics is responsible both for DM production and its decays.
As suggested in \cite{unified}, further experimental tests could rely on the flavour composition of primary neutrinos. 
It would be of course also interesting to explore the potential of future planned 100 TeV colliders 
and possible links with flavour anomalies. 

\vspace{-1mm}
\subsection*{Acknowledgments}

PDB acknowledges financial support from the STFC Consolidated Grant ST/T000775/1.
AM is supported by a DiscNET/NGCM scholarship. PDB wishes to thank the organisers of the 
workshop on {\em Neutrino Theories} held at the Institute for theoretical physics at UAM, Madrid,  
16 May - 17 June and the organisers of the workshop on the {\em Standard Model and beyond} held 
at the Corf\`{u} Summer Institute, 28 August-  8 September 2022, where part of this work was carried out.
During these workshops he could benefit from stimulating talks and discussions with Sasha Belyaev, Thomas Hambye, Steve King, 
Rocky Kolb, Gino Isidori, Nick Mavromatos, Apostolos Pilaftsis, Subir Sarkar, 
Mikhail Shaposhnikov, Alexey Smirnov and Jim Talbert.

\end{document}